\documentclass[12pt,preprint]{aastex}
\input epsf.tex
\begin{document}
% Prints a large "DRAFT" diagonally across each page
% Does not show up in TeXview
% \typeout{Prints "DRAFT" on each page; does not show in TeXView}
% \special{!userdict begin /bop-hook{gsave 200 30 translate
% 65 rotate /Times-Roman findfont 216 scalefont setfont
% 0 0 moveto 0.95 setgray (DRAFT) show grestore}def end}
\title{The Pulsar Planets: A Test Case of Terrestrial Planet Assembly}
\author{Brad M. S. Hansen\altaffilmark{1}, Hsin-Yi Shih\altaffilmark{1} \& Thayne Currie\altaffilmark{1,2}
}
\altaffiltext{1}{Department of Physics \& Astronomy, and Institute of Geophysics \& Planetary Physics, University of California Los Angeles, Los Angeles, CA 90095; hansen@astro.ucla.edu,jennywho@ucla.edu}
\altaffiltext{2}{Harvard-Smithsonian Center for Astrophysics, 60 Garden Street, Cambridge, MA 02140; tcurrie@cfa.harvard.edu}

%\slugcomment{\it 
%}

%\lefthead{Hansen {\it et al.}}
%\righthead{GC IMBH}

\begin{abstract}
We model the assembly of planets from planetary embryos under the conditions
suggested by various scenarios for the formation of the planetary system around
the millisecond pulsar B1257+12. We find that the most likely models fall at
the low angular momentum end of the proposed range. Models that invoke
supernova fallback produce such disks, although we find that a solar
composition disk produces a more likely evolution than one composed primarily of
heavy elements. Furthermore, we find that dust sedimentation must occur rapidly
as the disk cools, in order that the solid material be confined to a sufficiently
narrow range of radii. 
A quantitative comparison between the observations
and the best-fit models shows that the simulations can reproduce the observed eccentricities
and masses, but have difficulty reproducing the compactness of the pulsar planet system.
Finally, we examine the results of similar studies
of solar system terrestrial
planet accumulation and discuss what can be learned from the comparison.
\end{abstract}


\section{Introduction}

Although the discovery of the planet 51~Pegasi~b (Mayor \& Queloz 1995) is often
hailed as the start of the extrasolar planet revolution, the field of extrasolar 
planet studies began in 1991, with the discovery of the
two planet-sized bodies orbiting the pulsar PSR~1257+12 (Wolszczan \& Frail 1991). 
Subsequent study has confirmed this identification by identifying the effect 
of mutual gravitational
perturbations between the two planets in the timing residuals, thereby also
constraining (Wolszczan 1994; Wolszczan et al. 2000a;
Konacki \& Wolszczan 2003) the inclinations and planetary masses, the current values of which
are given in Table~\ref{Params}.
Yet, despite their early discovery, the origin of the pulsar planets have received little
attention after the initial burst of post-discovery papers (summarised in Podsiadlowski 1992).

Part of this neglect is no doubt due to the unusual nature of the host star and the
relative uniqueness of the planets' natal circumstances. However, as has been noted
before (Phinney \& Hansen 1993), the formation scenarios all produce a qualitatively similar outcome -- namely
a gaseous disk on a compact scale around a 1.4$M_{\odot}$ pulsar. Thus, the exact details of the
formation 
 are relevant primarily in the manner in which they determine
the size of the original mass and angular momentum budget, and also the disk composition. 
The subsequent evolution of the
gaseous and planetesimal disks are expected to follow a path very similar to the one they
would follow if orbiting a normal main sequence star of similar mass. As such, studying the
formation of the pulsar planets can offer us insights into the formation of terrestrial planets
as a whole. This will be of particular interest in the next few years as the searches for
planetary companions to main sequence stars continues to push down into the `super-earth' 
regime (e.g. Rivera et al. 2005; Udry et al. 2007; Mayor et al. 2008).

In that spirit, we wish to examine the quantitative evolution of pulsar protoplanetary disks
that arise from different formation scenarios. In Currie \& Hansen (2007), hereafter Paper~I,
we examined the expansion and evolution of gaseous disks for a range of mass and angular momentum,
 and identified the manner in which they laid down solid material
that might eventually coalesce to form planets. In this paper we wish to extend that analysis
to the question of how planetesimals, formed from such a distribution, would assemble into a
final planetary configuration, and how the resulting planetary systems compare to that observed.
In \S~\ref{Setup} we review the outcomes of Paper~I and how we construct the initial conditions
for the simulations described in this paper. In \S~\ref{Sims} we describe the qualitative properties of
the final planetary systems in each of the simulated scenarios. In \S~\ref{Comp} we compare these results to the observed system and discuss the issues of broader relevance in \S~\ref{Discuss}.

\section{Gaseous Disks}
\label{Setup}

Essentially all of the proposed pulsar planet provenances invoke the capture of material
from an external source, resulting in the formation of a gaseous disk of material in orbit
around the pulsar. The nature of the interaction that generates the material will determine
the mass and angular momentum of the initial disk, but all scenarios result in a disk that
is initially quite compact, with characteristic scales $\sim 1 R_{\odot}$. Thus the initial
state is one of super-Eddington accretion onto the pulsar from a hot disk. As mass is accreted
onto the star some material moves outwards to conserve angular momentum in the traditional 
 accretion disk manner. Although the physics that describes the evolution of the outer edge of this expanding
 disk is similar to that of normal protoplanetary disks, it offers an interesting variation on the
traditional situation, in which the scale of the disk is set by the infall of material from larger
radii. As we explain below, this leads to important differences that may help to explain the observed configuration.

In paper~I we describe the evolution of such disks for a range of total angular momentum. We motivate this 
by making reference to two specific scenarios, although they would apply equally well to different situations
that produce disks with the same global properties.
The first scenario is the formation of a disk that results from supernova fallback material. Models of this
kind lie at the lower end of the range of the disk angular momentum distribution. The second scenario
results in the formation of a disk by the tidal disruption of a companion star with which a newly
kicked neutron star collides. This yields angular momentum $\sim 10^{51}$--$10^{52}$ ergs s, which
lies at the upper end of the theoretical distribution. In paper~I we incorporate a model to
describe the thermal and viscous evolution of the disk, assuming a traditional metallicity for
the gaseous material (which is consistent with most, although not all, formation scenarios). We
also consider two variations for the viscous evolution. In the first case, we simply assume a 
traditional parameterised $\alpha$-viscosity (Shakura \& Sunyaev 1973). In the second case, motivated by recent
considerations regarding the ultimate physical origin of the $\alpha$ viscosity, we consider a
two-viscosity model in the spirit of the `layered accretion' model of Gammie (1996). In this
model, we assume that the viscosity goes to zero when the local temperature of
the disk  material at a given radius drops to a level where the material is no longer expected
to be ionized. This has the effect of regulating the outwards spread of the disk (Hansen 2000; paper I) and imposing a final mass profile that depends on the total angular momentum of the disk.

Into each of these evolutionary models we also incorporate a prescription for the deposition of
solid material, which is assumed to decouple from the gas and sediment to the midplane when the
temperature drops below a certain threshold. For the purposes of this paper, we will assume that
the local deposition of solid material determines the initial distribution of planetesimal mass. There
are a variety of proposed physical effects, such as radial drift due to gas drag or `Type~I migration', that
may affect this distribution, but the degree to which such processes operate in reality is poorly
understood and a matter of discussion. We have opted to calculate the simplest model, and will discuss
possible refinements in \S~\ref{OtherStuff}.

% We opt instead to simply
%consider the mass profile to be set by the gas disk, and allow for an overall `efficiency factor',
%with which we can reduce the amount of mass in the planetesimal disk relative to the original
%deposited solid material. {\em rephrase some of this}

Using this model, we can predict the form of the solid material profile at the end of the
gaseous evolution and adopt this as the initial conditions for the planetesimal assembly
simulations.

\section{Planetesimal Disks}
\label{Sims}

The models in Paper~I provide a surface density profile of solids that provide the
initial condition for the next stage of the evolution, namely the assembly of the
rocky material into the observed planets. In the standard model of planetary assembly, the
 initial stage from planetesimals
to planetary embryos is essentially a local phenomenon, and proceeds until most of
the material is concentrated in a series of `oligarchs' (e.g. Kokubo \& Ida 1998),
 which contain the majority
of the mass within an annulus whose width is roughly several times the Hill sphere of the oligarch
in question. The final stage of planet formation is then the subsequent collisional
assembly of this population of planetary embryos into the final planets. We wish
to simulate
this final stage evolution of the planetesimal disks in each case
discussed above, using the planetary dynamics code {\tt Mercury6} (Chambers 1999). 
This code incorporates a hybrid scheme which treats the longer-range interactions
using a symplectic integrator but treats close encounters with a direct Bulirsch-Stoer integration. This makes it
particularly useful for our purposes, because the transition from collisional accumulation
to a dynamically quiescent configuration is presumably what will determine the final
planetary configuration.

We will assume an initial distribution composed of equal mass planetary embryos, whose
spatial distribution is chosen to mimic the solid material surface density profile left
at the end of gas disk evolution as described in Paper~I, for each individual scenario. We choose to start with these
masses rather than the masses estimated for the final stage of oligarchic growth because
the surface densities we use below are often significantly larger than those normally
used in the solar system context, and sometimes lead to estimated oligarchic masses $> 1 M_{\oplus}$.
 In such cases, it is probably not a good approximation to assume the assembly proceeds to 
full oligarchic completion without interacting with neighbouring annuli. In effect, 
the distinction between the local runaway accretion into oligarchs and the final completion of assembly
 is blurred. 
We therefore
choose to start with smaller masses and more bodies in order to be sure not to miss some of the
essential dynamics. This amounts to essentially starting the simulation close to the end of the oligarchic
growth phase (we begin at the point where is still a significant population of small bodies) but before the individual
annuli are fully cleared.

Let us now consider each of the cases in turn. We consider bodies with densities $1.5 g.cm^{-3}$,
initially on perfectly circular, coplanar orbits, 
and assume an integration accuracy of $\epsilon = 10^{-11}$ per timestep (which is taken to
be 4 days,
 so that all relevant orbits are well resolved). We assume all physical
collisions result in mergers, although for some of the more compact disks, encounter velocities 
are high enough that fragmentation is a real possibility.

\subsection{Large Disks: Fully Viscous Evolution}
\label{TDFV}

Disks with total angular momentum $\sim 10^{52}$ ergs\, s lie at the upper end of the distribution
produced in the various proposed scenarios. One scenario that gives rise to such a disk is
the
tidal disruption of a close stellar companion to the progenitor star, which 
 is disrupted by the passage of the post-supernova kicked neutron star (e.g. Phinney \& Hansen 1993).
The large angular momentum means that the disk expands to radii $> 1$AU before becoming cool enough
to form dust and allow the deposition of solid material. If the disk remains fully viscous 
throughout, the resulting solid material is deposited in a relatively flat surface density profile and extending out as far as 7~AU.
Following the results of paper~I, we assume the initial surface density of solids is constant,
with a value $\Sigma = 10 g/cm^{3}$, out to 7~AU. The mass is spread initially between
100 identical bodies, each of mass 0.58 $M_{\oplus}$.
 This results in a planetesimal disk of total
mass $M_0 = 58 M_{\oplus}$.

Figure~\ref{Snap1} shows the evolution of the resulting planetary embryo swarm. As
some of the bodies grow to  masses $> 1 M_{\oplus}$, the eccentricities and
inclinations are amplified by the mutual scattering. On timescales $\sim 10^7$ years,
the outer edge of the swarm is pushed out to $\sim 20$~AU, and there are five bodies
with mass $> 3 M_{\oplus}$. After $10^8$ years, there is a dominant body, with
mass $16.3 M_{\oplus}$, semi-major axis 0.91~AU and an eccentricity of 0.263. There
is also still a significant amount of mass at larger radii. Figure~\ref{Massive1} shows
the radial evolution of the body that grew to be the most massive one. It initially
began at $\sim 3$~AU, and underwent a brownian random walk during the initial accumulation,
 migrating out as far as 8~AU before undergoing an inward
migration as its mass began to grow. Once the mass is large enough to scatter smaller
bodies outwards instead of accreting then, it begins to preferentially lose binding energy,
 much in the same manner as happens in
our solar system (e.g. Fernandez \& Ip 1984). The fact that the available reservoir of material
is comparable in total mass to the scattering body means that this system suffers from the same
kind of instability as 
 described by Murray et al. (1998), albeit in a slightly different context. The rapidity of the migration also explains why some of
the material is left in partially scattered orbits (Hansen 2000), even though the dominant body is now
capable of ejecting such material from the system (see \S~\ref{Dynamics}). In fact, only 1.3 $M_{\oplus}$ of material is ejected during the evolution.

The size of the dominant remaining body also invites speculation as to whether it could form
a Jupiter-class gas giant planet, since the mass is larger than the $\sim 10 M_{\oplus}$ threshold
for nebular accretion in the context of the core accretion scenario for giant planet formation
(e.g. Mizuno et al. 1978; Pollack et al. 1996). In Figure~\ref{Massive1} the dotted lines show the points at which the
body crosses the mass thresholds of $5 M_{\oplus}$ and $10 M_{\oplus}$. These
occur on timescales $30$Myr \& $80$Myr respectively, suggesting that giant planet formation is
unlikely, since the gas density drops rapidly on Myr timescales (Paper~I). Thus, the final
configuration is likely to be accurately reflected in Figure~\ref{Snap1}.

The final configuration from this model is clearly wildly discrepant with
regards to the observed system PSR B1257+12, as the observed planets are smaller
and much closer to the pulsar than anything produced in this model. This is the case
in every model realisation of this scenario, as can be seen in Table~\ref{CompTab}.
 Indeed, the dynamical
signatures that result from this type of system would produce time-of-arrival fluctuations
of order 10~ms or more, and are thus potentially detectable around even normal pulsars.
The larger orbits also imply orbital periods that are frequently several years, so that
it is an interesting question as to whether any objects of this ilk are lurking in the
timing noise of presently known systems.

\subsection{Large Disks: Layered Accretion}
\label{TFLA}

The model of the previous section clearly produces too much mass on too large
a scale to produce anything resembling the observed pulsar planets.
However, a fully viscous disk is unlikely to be an accurate model for any protoplanetary
disk on these scales. In paper~I, we also considered an evolutionary sequence for a large disk within the
context of the
 layered accretion model of Gammie (1996). In this case,
 the final solid surface density is still approximately constant
with radius, but with a higher value ($\Sigma = 20 g/cm^{3}$), and the disk is more compact 
(with an outer edge of 4~AU). Once again, we begin with 100 bodies, this time each with mass
$0.38 M_{\oplus}$.
 The total mass in this case is thus $38 M_{\oplus}$.

Figure ~\ref{Snap2} shows the corresponding system configurations at 1, 10 and 100~Myr. The
behaviour is similar to that observed in Figure~\ref{Snap1}, except that the evolution proceeds
a little faster because of the more compact configuration. Once again, the final state is
a moderately massive body (in this case 12 $M_{\oplus}$), with semi-major axis 1.2~AU and
eccentricity 0.206. In this case, at 100 Myr, this is only the second innermost body, with
another, 2.3 $M_{\oplus}$, body at smaller radii, pushed inwards as the larger body migrated.

In both this case and the former, 
the dynamical evolution of this system of planetary embryos results 
 in  a single body which rapidly accretes a substantial
fraction of the total mass, but probably on a long enough timescale that the gas inventory
was reduced before it had a chance to form a gas giant planet through core accretion.
This is a consequence of the excretion disk nature of the protoplanetary disk in this
context, which results in a smaller and more rapidly evolving gas budget than in a traditional
protoplanetary disk.

\subsection{Small Disks: Fully Viscous}

At the lower end of the proposed angular momentum range is a disk with total angular momentum $\sim 10^{49}$ ergs\, s.
A simple model which produces such is disk is the supernova 
 fallback disk model, in which some fraction of the material in the progenitor core remains
bound even after the passage of the supernova shock wave, and falls back to form a disk around
the newly born neutron star. Disks that result from such a scenario have masses $<0.1 M_{\odot}$
and total angular momentum $\sim 10^{49}$ ergs\, s (Chevalier et al. 1989; Hashimoto et al. 1989;
Lin et al. 1991; Menou et al. 2001). The amount of material that actually expands to the scales
of interest is smaller, $\sim 10^{-3} M_{\odot}$. This also depends on the composition of the material. If the fallback material comes from the outer layers of the star, it may contain significant
Hydrogen and Helium, and may evolve in a manner similar to a solar composition disk. On the other
hand, if the material originates from the interior, onion-skin layers of the star, the evolution
may be quite different because of the heavy element-dominated composition. In this section we
will discuss the solar composition disk, leaving the heavy element disk to \S~\ref{Kristen}.

Once again, the rate of expansion is also affected by what one assumes for the viscous evolution of the disk.
In the case of a fully viscous disk, with $\alpha \sim 0.01$, the final distribution of solid
material can be characterised by a surface density profile 
\begin{equation}
\Sigma = 30 g cm^{-2} \left( R/ 1 AU \right)^{-1}
\end{equation}
with a cutoff at 1.5~AU. The resulting mass is lower than in the previous models, only 10.6$M_{\oplus}$. 
Once again, we spread this between 100 initial bodies, each of mass 0.106~$M_{\oplus}$.

The dynamical evolution is shown in Figure~\ref{Snap3}. 
The comparison with observations is much more encouraging in this case,
 primarily
because the mass budget is much closer to that found in the observed system.
The agreement is not perfect however. Different realisations of this scenario (Table~\ref{CompTab}) all
produce several planets with $> M_{\oplus}$, but spread over a range that extends
beyond 1~AU, which is markedly more extended than the observed system. In the
case shown here, we have four planets, with masses $1.4 M_{\oplus}$, $2.2 M_{\oplus}$,
$2.2 M_{\oplus}$ and $3.4 M_{\oplus}$, with semi-major axes extending from 0.18~AU to
1.2~AU. There is also potential for smaller, dynamically decoupled bodies at $> 2$~AU.

\subsection{Small Disks: Layered Accretion}
\label{LayeredFall}

However, as we noted before, a fully viscous evolution is unlikely and a more physically motivated
gaseous evolution is provided by the layered disk model.
In the case of layered accretion, the remaining solid surface density is even steeper, with
\begin{equation}
\Sigma = 8 g cm^{-2} \left( R/ 1 AU \right)^{-5/2}
\end{equation}
with a cutoff at 1~AU. Once again, the simulations are begun with 100 bodies, with individual
masses~0.082 $M_{\oplus}$. The resulting total mass is 8.2$M_{\oplus}$.

The results of this scenario are shown in Figure~\ref{Snap4} and Table~\ref{CompTab}. Again, there are some qualitative
agreements with the observations. Planets of approximately the right mass are formed in the
region 0.3--0.5~AU. However, as in the previous case, our final configuration contains several
more planets than are observed. In this case we are left with five planets, all of whose masses
are $>0.5 M_{\oplus}$ and are thus readily observable. We have also verified that this system
is dynamically stable by letting it run out to $10^9$~years. 

\subsection{Heavy Element Disks}
\label{Kristen}

Although we used the supernova fallback scenario to motivate our choice of total angular momentum
above, we still described the evolution of the gaseous disk in terms appropriate to cosmic composition.
In many scenarios, the remaining disk is hydrogen poor and dominated by heavier elements, such as Carbon
and Oxygen. In such cases, the
higher ionization potentials of the dominant disk material means that the 
disk is likely to become neutral at much higher temperatures and thus remain quite compact (Chevalier 1989,
Hansen 2000). Menou, Perna \& Hernquist (2001) discuss the evolution of such gaseous disks in some detail. They
find that the expanding disk is likely to stall (because heavy elements become neutral at much higher temperatures)
 while still very compact, indeed at radii smaller than
the tidal disruption radius ($\sim 10^{11}$cm) for a planet-size rocky body. In this event, the starting
point for a planetary embryo calculation should thus be a very compact disk, which could potentially scatter
bodies out to larger radii, where they might form pulsar planets as a residual, scattered population.

To model this situation, we start with  an annulus of width 0.05~AU, located at 0.1~AU. If we imagine that
the planets move outwards to conserve angular momentum while most of the mass goes inwards, in the traditional
fashion of accretion disks, then we need to assume an initial disk mass that is larger, $\sim 20 M_{\oplus}$.
However, the end result of the simulation does not contain any planets at radii significantly larger than
the original annulus, but rather the original inventory is spread amongst three larger planets,
spread between 0.049~AU and 0.112~AU. The problem is essentially that the planets involved are not big
enough to drive an outwards migration in the manner of our outer solar system (e.g. Fernandez \& Ip 1984;
Hahn \& Malhotra 1999), because the repulsive force is mediated through multiple scatterings of smaller bodies.
When the scatterers are $\sim M_{\oplus}$, this leads to collision and accretion more frequently, short-circuiting
any tendency towards outwards migration (see \S~\ref{Dynamics}). We repeated this calculation with several realisations, and found no significant transport of mass outwards in any of them.
 For completeness, we also performed the same
simulation using an initial disk inventory of $10 M_{\oplus}$, initially spread amongst 100 bodies, which resulted in two planets, of mass 4.9$M_{\oplus}$
and 5.1$M_{\oplus}$, located at 0.043~AU and 0.088~AU. Thus, it appears as though an initially compact planetesimal disk
cannot produce pulsar planets on the scales observed. We note that the accumulation of
planets in this scenario is extremely rapid, with the final system essentially in place within $10^3$~years,
although the accumulation may be lengthened if collisional erosion is considered, as the encounter velocities
in such a compact disk are likely to be higher than in the other cases.

\section{Do Any of These Models fit?}
\label{Comp}

%Things to consider -- planet masses and seperations; timescale for accumulation; existence and
%survival of bodies on larger scales; resonant interactions.

A qualitative comparison between the observations and the data suggests that the layered disk evolution
of the smaller disk represents the most likely match to the observed pulsar planets. The larger disks
simply produce too much mass on too large a scale, resulting in planets that are far too massive.
In the case of the smaller disk, the final masses are closer to those observed, although spread over too large a
range in radius if the disk is assumed to evolve via the $\alpha$ disk model. In the case of the heavy element
disks, one can reproduce the number and masses of the planets, but they are invariably found too close to the central star.
However, even in the case of our favoured scenario, the comparison is not perfect. Multiple realisations of the
small, layered disk all result in more planets than those observed -- essentially, nature incorporates the same
amount of mass into fewer bodies than our model predicts.

In order to make the comparison more quantitative, Table~\ref{CompTab} shows the values
of a variety of statistics proposed by Chambers (2001) for the quantification of terrestrial planet
systems, as applied to simulations of the above
scenarios. These include $S_m$, the fraction of the total mass stored in the largest object, $S_s$, a
statistic that corresponds roughly to the mean spacing between planets in terms of their Hill radii, 
$S_d$, the normalised angular momentum deficit (a measure of how circular the orbits are) and
$S_c$, a mass-concentration statistic, which measures how localised in radius the planets are.
We also include the mass-weighted semi-major axis, $S_a$, of Chambers \& Cassen (2002). We
 restrict our attention to those particles in the simulation which would yield fluctuations in
the timing residuals $> 4 \mu s$ and with periods less than 20~years. This is in order to approximately
match the observational limits. Finally, we also show the value of the most massive planet ($M_{big}$)
and it's location ($a_{big}$).

The numbers bear out the qualitative observations above. Particularly striking is the disparity between
the observed and simulated values of $S_c$, the localisation statistic. The observed pulsar planets have
a value of $S_c$ an order of magnitude larger than the results from all the simulated cases,
indicating that most proposals produce systems that are simply too spread out in radius. Note that the
pulsar planets are even more severely localised (by a factor of 4) than our own terrestrial planet system.
The simulation planets are also, in general, more eccentric than the observed planets, as indicated by
the $S_d$ statistic\footnote{This is a common problem with simulations of our own solar system formation
as well (e.g. Chambers 2001).}. Figure~\ref{SS2} shows this result in graphical terms. Figure~\ref{MA0} shows the
comparison of models and observations in terms of $M_{big}$ and $S_a$, which reflects whether the final
planets are of the right size and in the right general location. We see here the basis of our earlier assertion
that the smaller disk models provide the best qualitative match.

In performing the above comparisons, we have assumed that the current census (Konacki \& Wolszczan 2003)
 of the pulsar planet system
(which stands now at three objects) is
complete. There have been reports in the past (Wolszczan et al. 2000b) of additional timing residuals
in the B1257+12 data that might be indicative of the presence of additional small planets at larger radii.
However, subsequent observations suggest these are due to effects related to the propagation of radio waves through
the interstellar plasma, and not gravitational in origin (Alex Wolszczan, private communication). We have tested
the sensitivity of the above measures to the presence of additional small bodies whose contribution to timing
residuals would have amplitudes $< 1 \mu s$, and found negligible influence. 

One statistic that is affected by observational uncertainties is the $S_d$ statistic -- the angular momentum
deficit. The relative inclination of the two large planets reported by Konacki \& Wolszczan is $6^{\circ} \pm 4^{\circ}$. This has a significant effect on $S_d$ because the statistic is sensitive to any deviations from circular, coplanar
orbits. Thus, in Figure~\ref{SS2} we show the error in $S_d$ due to the $1\sigma$ error in relative inclinations.

\subsection{Dynamics}
\label{Dynamics}

The results of the preceding sections can be understood within the context of the
theory of planetary assembly as developed for our own solar system. Goldreich, Lithwick \& Sari (2004)
recently reviewed the understanding of the final stages of planetary assembly, with specific 
focus on the origin of Uranus \& Neptune. Within the terminology of Goldreich et al., our simulations
apply to the period referred to as `completion', when all small bodies have been consumed and the
dynamics is governed by the interaction and merging of bodies of planetary embryo size. 

There has been a long-standing discussion as to whether there is a need for a 
dynamically cooled 
population of small bodies (not present in the simulations shown here) during these late stages of planetary accretion.
The argument in favour of this population is that, without it, the accumulation of material to form Uranus
and Neptune is too slow -- the result of the fact that the rate of accretion is reduced as the velocity dispersion
of the accreting bodies is increased by scattering off the growing planet\footnote{Although this problem may be
alleviated if Uranus and Neptune formed closer in and migrated outwards -- Thommes, Duncan \& Levison 1999.}. These considerations do not, in fact,
impact the simulations shown here because the systems we simulate are more compact, and the planets can accumulate
within a finite amount of time without any additional dynamical cooling. As an illustration of this, we can recast
equation~(56) of Goldreich et al. (2004), which describes the characteristic accretion time for a large planet, in the form 
\begin{equation}
T_{\rm form} = 15 {\rm Myr} \left( \frac{a}{1 AU} \right)^3 \left( \frac{u}{v_{esc}} \right)^2
\label{GLS}
\end{equation}
where $u$ and $v_{esc}$ are is the characteristic random velocity of the accreting bodies and $v_{esc}$ is the
escape velocity from the surface of the growing planet. Our simulations, without any dynamical cooling, automatically
produce $u \sim v_{esc}$. The estimated formation time of $\sim 15 \rm Myr$ is in good agreement with simulations shown
in Figures~\ref{Snap3} and \ref{Snap4}, for example.

Application of this formula to the simulations shown in Figures~\ref{Snap1} and \ref{Snap2} also explains why it
takes somewhat longer to form the largest body in those systems. The accumulation time is longer because the bodies
start forming at 2--3~AU, and so $T_{form}$ can be an order of magnitude larger. One aspect of these simulations
not described in the Goldreich et al. treatment is the inwards migration of the dominant body. For this, we turn to
 the framework laid out
by Tremaine (1993). As the principal body grows in size, it scatters the smaller bodies in the vicinity. The
initial evolution can be characterised as a growth of the velocity dispersion, as described by Goldreich et al.,
but as the growth continues, a more accurate characterisation is 
 as a diffusion process in energy space, with fixed periastron. It is this process that is thought to drive the
evolution of comets towards the Oort cloud (e.g. Fernandez \& Ip 1984; Duncan, Quinn \& Tremaine 1987).
 Using a diffusion coefficient based on numerical
simulations, Tremaine derives a criterion for a mass threshold above which a planet can drive orbital evolution
of smaller bodies on a characteristic timescale $T$
\begin{equation}
 M > 13 M_{\oplus} \left( \frac{T}{\rm 10 Myr} \right)^{-1/2} \left( \frac{a}{1 AU} \right)^{3/4}
\label{T93}
\end{equation}
where we have assumed a pulsar mass of 1.4$M_{\odot}$ and where $a$ is the semi-major axis of the
scattering planet. Thus, combining equations~(\ref{GLS}) \& (\ref{T93}), we may derive a criterion for a
planet to migrate during it's accumulation process
\begin{equation}
M > 11 M_{\oplus} \left( \frac{a}{1 AU} \right)^{-3/4},
\end{equation}
which corresponds quite well to the observed evolution.

Once a planet has begun to scatter other bodies in the system, the question is whether the cumulative
effect of encounters and scatterings is to accrete material or eject it from the system. Once again,
Tremaine estimates the threshold mass for this process to be 
\begin{equation}
 M_p \geq 9 M_{\oplus} \left( \frac{a}{1 AU} \right)^{-3/2}
 \left( \frac{M_*}{1.4 M_{\odot}} \right)^{3/2} \left( \frac{\Delta i}{30^{\circ}} \right)^{-3/4}
\end{equation}
where $\Delta i$ indicates the average range of inclinations of the population of small
bodies being scattered. This number is taken from the simulations directly. If the vertical
velocity dispersion $\sigma_z$ is generated by scattering off the same body (mass $M_p$) as is driving the collisional
evolution, then 
\begin{equation}
\Delta i \sim \sigma_z/r \Omega \sim \frac{1}{3} \frac{M_p}{M_*} \frac{a}{R_p}
\end{equation}
where $R_p$ is the radius of the planet. Assuming a mean planet density of $3 g/cm^{-3}$, this
results in a modified collision/ejection transition mass
\begin{equation}
 M_p \geq 16 M_{\oplus} \left( \frac{a}{1 AU} \right)^{-5/2}
 \left( \frac{M_*}{1.4 M_{\odot}} \right)^{5/2}.
\end{equation}

Figure~\ref{MA3} shows these criteria, along with the final states of two simulations,
those shown in Figures~\ref{Snap2} and \ref{Snap4}. Bodies that lie above the dotted line
labelled `Migrate' will migrate inwards faster than they can accrete. Bodies that lie below
the dotted line labelled `Accrete' will preferentially accrete the bodies they encounter,
while those above the line will eventually eject bodies through repeated scatterings. We
see that the small disk simulation (solid points) is firmly within the regime where all mass
is eventually accreted into the largest bodies,
 while the large disk simulations (open points)
are capable of losing material to ejection, and large bodies are able to migrate a
significant amount. The largest body in this latest simulation is located
near the intersection of the two criteria, which is not a coincidence. The body grows to the
point where it migrates inwards by scattering companions, but then eventually stalls as a result
of reaching the threshold between accretion and ejection. Thus, there is a qualitative difference
in the two cases -- small disks accumulate material together quasi-locally, while the final configuration
of a larger disk is regulated by global considerations of energy and angular momentum.
 This is, to a large extent, also the difference
between the inner and outer parts of our own solar system.

For completeness, we note that the presence of a population of smaller and dynamically colder bodies (neglected
in these simulations)
in \S~\ref{TDFV} or \ref{TFLA} could potentially speed up the accretion so that a large core is formed
before the gas disk dissipates (in a similar spirit to the models of our own solar system). In that
event, one could possibly form a true gas giant planet in those scenarios. However, in the absence of
any observational evidence, we have not considered that model in detail.

\section{A refinement to the scenario}

The models presented in \S~\ref{Sims} produce an encouraging, but not entirely
satisfactory result. A disk of total angular momentum $\sim 10^{49}$~ergs\, s, 
made of approximately solar composition, and whose gaseous evolution proceeds
via the layered disk model, can produce planets of approximately the correct
mass, in approximately the correct location, to be a plausible progenitor of
the PSR B1257+12 system. However, the models generically produce planets that
are too numerous, too spread out and more eccentric than the observations. This
suggests that the model, as described, does not yet tell the full story.

\subsection{Rapid Sedimentation}
\label{Dump}

In examining the results of paper~I, we can identify a probable reason why the mass distribution
is more extended than the observations indicate. 
 The initial deposition of mass is limited
to the region 0.4--0.6~AU, as the outer edge of the expanding disk drops below the temperature
threshold for dust particles to condense out.
As the disk
continues to evolve, more mass expands outwards causing the surface density at the outer edge to increase, which, 
in turn, 
results in an increase in the local temperature. If one assumes, as we did in Paper~I, that the sedimented material
is reabsorbed into the gas when the local temperature increases above condensation temperature, then the deposition
profile will be influenced by the continued evolution of the gaseous disk and the final profile will only be layed
down at late times, when the disk has ceased to expand significantly.
 The lower angular momentum disk scenarios
are more sensitive to this issue than larger, more extended disks, such as a normal protoplanetary
disk or the disks that result from tidal disruption. However, if
 the dust that sediments out forms
larger and more robust agglomerates on a timescale short compared to the disk
evolution time, then it will not be reabsorbed into the reheated gas, and then
 the correct distribution
of solids is more accurately represented by the
initial deposition profile, such as seen in Figure~4 of Paper~I.

Thus, we have also performed simulations in which the mass is distributed in a uniform surface
density annulus between 0.4 and 0.6~AU. For a disk mass of $9 M_{\oplus}$, this results in a
surface density $\sim 380 g/cm^{-2}$. We simulate this as before, starting with 300 bodies, each
of mass $0.026 M_{\oplus}$.\footnote{Three simulations were also performed with the same total
mass spread amongst 800 bodies, with quantitatively similar results.}
The evolution of a planetary embryo swarm drawn from such a starting disk is shown in Figure~\ref{Snap5}. 
The evolution is
particularly rapid in this case, being largely complete within $10^6$~years. The more rapid pace
is simply because the material is concentrated in a limited range of radii. We also note that there
is relatively little orbital evolution of the largest bodies. The eventual outcome is also potentially
the most encouraging of all the calculations, as the number, masses and locations of the bodies
are now qualitatively similar to those observed. 

Figure~\ref{SS3} and Table~\ref{CompTab2} show the quantitative comparison. This is much better than 
that shown in Table~\ref{CompTab} or Figure~\ref{SS2}. The values of $S_d$ and $S_c$ are closer to the observed system than before (and comparable
to the terrestrial system, although $S_c$ is still not quite large enough to match the pulsar planets).
 Other measures are also
in approximate agreement -- the masses of the planets are similar, and the mass-weighted semi-major
axis $S_a$ is closer to the observed value, although still approximately 25\% too large.

Although the values of $S_a$ are too large, we note that there is little dispersion from one realisation
to the next, with $S_a$ almost always lying close to the midpoint of the annulus.
 Thus, given the encouraging agreement in other parameters, we might improve the fit 
 by simply
postulating an annulus at a given radius -- effectively allowing for the possibility that our gaseous evolution
models might be slightly overestimating the outward expansion of the protoplanetary disk.
 Thus, we have calculated similar models but centered at 0.4~AU rather than 0.5~AU.
We have also narrowed the annulus to investigate whether this can help with the mismatch in $S_c$.
The second set of results in Table~\ref{CompTab2} thus show the results of an annulus that starts
 with material between 0.35--0.45~AU. The results are indeed qualitatively similar as before, but now
 with $S_a$ matching the observed value, as expected. The simulations can easily match the observed
$S_d$ and get within a factor of two of $S_c$, in extreme cases. The values of $S_a$ and $S_m$ also
match well, and are shown in Figure~\ref{SS4}.
Thus, viewed overall, the simulations now provide plausible matches to the observations 
in all the measures proposed by Chambers, with the possible exception of $S_c$.

\section{Discussion}
\label{Discuss}

The ultimate goal of this paper is to evaluate the various proposed models for pulsar planet formation
within the context of a physically motivated theory of protoplanetary disk evolution and planetary
assembly. A secondary goal is to use our understanding of the successful model(s) to illuminate our
understanding of terrestrial planet formation in general, including in our own solar system and
other extrasolar planetary systems.

\subsection{Evaluating the Formation scenarios}

We have essentially considered formation scenarios to fall into one of two classes -- either
solar composition protoplanetary disks, which evolve in a fashion similar to those around normal
stars, or disks composed primarily of heavy metals, whose evolution we assume to be truncated at small 
radii according to the models of Menou et al. (2001).

 In the former case, these models form a one-parameter
family, whose evolution is dictated by the overall angular momentum. We have examined the consequences
of planet formation in this family of models and find that those with large values ($10^{51}$--$10^{52}$ ergs\, s)
produce too much solid material over too large a range of scales. The collisional assembly of the solid material
from these models yields planets that are too large and with a significant number of secondary planets, whose
presence would be easy to detect, possibly even around normal pulsars. The models which produce planetary
systems that qualitatively resemble the observed system have total angular momenta at the low end of the
proposed range ($\sim 10^{49}$ ergs\, s). One proposed scenario that produces such values is the supernova
fallback scenario. However, in most versions of this scenario the composition of the material is dominated
by heavier elements, and the above evolution was calculated with an opacity and equation of state appropriate
to a  solar composition.

To examine the consequences of heavy element disks, we also calculated models that begin with planetary embryo
swarms localised at $\sim 0.1$AU, to see whether sufficient mass could be pushed outwards to the locations of the
pulsar planets, in the same manner as Uranus \& Neptune could have been forced outwards in our own solar system.
However, in this case, the masses are smaller and the material is deeper in the potential well, with the result that
there is very little expansion and the resulting planetary systems are far too compact.

The fact that our models favour the supernova fallback model is consistent with another analysis by Miller \& Hamilton (2001). They conclude that the B1257+12 pulsar is actually a relatively young pulsar, more closely associated (at least in
terms of origin) with normal, longer period pulsars than with the dynamically older population of millisecond pulsars into
whose company one would nominally place it based on the observed spin period. In part, their analysis is based on the
survival of the protoplanetary disk in the face of irradiation from the central object. This is an effect that we have
not considered in detail, based on the fact that the initial accretion rates from the disks we postulate are
 super-Eddington, so that the expanding disk is already ionized, and furthermore, the inner parts of the disk are expected to shield the outer disk from much of the central
object irradiation. Nevertheless, our models are in complete accord, as the disk masses we use fall well above the
threshold they require, and our models produce recognisable planetary systems on timescales $\sim 10^7$~years, appropriate
for the dynamical age of a newly born neutron star.

There is also independent observational evidence in favour of the existence of supernova fallback disks. 
Wang, Chakrabarty \& Kaplan (2006) report the presence of an infra-red excess associated with the isolated
young neutron star 4U~0142+61. This is interpreted as the reprocessing of X-rays by a source of dust in
orbit around the neutron star. The Wang et al. Spitzer measurements place a lower limit on the disk
outer edge $\sim 0.1$AU, consistent with our models but not significantly constraining. However, one should note 
that this source is a slowly rotating neutron star, rather than a rapidly rotating one like B1257+12, so
that this system cannot be considered an exact analogue at an earlier stage but only an example that such
phenomena do occur.

\subsection{Planet Assembly}

Within the context of the rapid sedimentation models in \S~\ref{Dump}, planet assembly occurs quite
rapidly. Figure~\ref{Assemble} shows the assembly and orbital histories of the two massive surviving
bodies in Figure~\ref{Snap5}. We see that the bulk of the mass growth occurs between 0.1 and 1 Myr,
with a significant fraction resulting from mergers with other objects of significant mass. On timescales
$>$1 Myr, the growth is slower, consisting essentially of the accretion of the remaining small bodies
in the vicinity. We see also the orbital evolution, with the repulsive interaction of planetary scattering
driving the two bodies to opposite ends of the original annulus.

Note that, despite the fact that our final model represents the outcome of a narrowly confined
initial distribution, we never end up with a single planet. This is presumably because scattering of planetesimals
can mediate a repulsive interaction between two accreting bodies, driving them apart. Formation of multiple final 
bodies also allows for simultaneous conservation of total energy and angular momentum in this largely
conservative system (e.g. Wetherill 1990).
 The most common outcome is two or three planets
of comparable mass. This is frequently the end result, but sometimes the process of accumulation does not
extend fully to completion.
This can be seen from the final panel in Figure~\ref{Snap5}, as the final system does contain two
 scattered bodies on orbits with semimajor axes between 1 and 2~AU. These appear dynamically decoupled from
the larger bodies and can thus potentially survive long enough to be observed. The total mass in this
scattered population is $\sim 10^{-3}$ of the total system mass, i.e. of order a moon's mass or less, at
an age of 100~Myr. This should yield timing residuals $\sim 10 \mu s$, which is potentially observable.
 At an age of 10~Myr (probably the youngest plausible age for such a pulsar), the mass
in this population is larger, $\sim 0.2 M_{\oplus}$, spread amongst several objects. 

Figure~\ref{Ast232} shows how such remnant objects become dynamically decoupled from their parent population.
Beginning at $\sim 8$~Myr, the periastron of the (eventual) outermost object in Figure~\ref{Snap5} starts to
move out slowly, so that the body no longer re-enters the region (delineated by dashed lines) where it is
scattered by the large bodies. This drift is driven by the interaction with another scattered body whose semi-major
axis is larger but whose periastron is smaller. The gravitational interaction between these two small scattered
objects continues until 13.5~Myr (shown as the dotted line), at which point the other body is ejected from the 
system entirely (recall the argument in \S~\ref{Dynamics} is only statistical -- a few bodies are ejected rather
than accreted), leaving a remnant body which is now dynamically decoupled from the main planetary region even though
it was formed there originally.

\subsection{Comparison with the Solar System}

One of the most interesting aspects of the pulsar planet system is that it represents an opportunity
to test the theory of terrestrial planet formation on an independent system. So, how does our preceding
discussion fit in with current notions about the assembly of the solar system terrestrial planets?

The solar system planets are usually assumed to originate from a planetesimal distribution drawn
originally from some variation on the minimum mass solar nebula (e.g. Weidenshilling 1977) or from a
viscous accretion disk model of the protoplanetary nebula (e.g. Bell et al. 1997). The lack of
remaining material in the region that encloses Mars and the asteroid belt is usually associated
with the disruptive influence of resonant perturbations associated with the giant planets (Wetherill 1992;
Lecar \& Franklin 1997; Nagasawa et al. 2000). Simulations of the final (completion) stages of accumulation 
from embryos into planets (Chambers \& Wetherill 1998; Agnor et al. 1999; Chambers 2001),
 based on such models have both successes and failures, often in common with our results in \S~\ref{Sims}.
For reasonable choices of parameters they produce plausible analogues of Earth and Venus, both in terms
of mass and position, although they consistently fail to reproduce Mercury and Mars accurately (when they produce
planets in the correct location, the planets are usually too massive). They also tend to produce systems that
are too eccentric and not sufficiently localised (Chambers 2001), in much the same manner as shown in
Figure~\ref{SS2}. A proposed solution for this latter problem requires a dissipation mechanism, in order to reduce the
epicyclic component of the surviving planets velocities. 

One possibility is that the final stages of accumulation occur while there is still a significant mass contained
in smaller bodies, which damp the eccentricity growth through dynamical friction (Goldreich et al. 2004;  O'Brien, Morbidelli \& Levison 2006). Whether such a significant small body population can last for long enough is a matter of some
debate (Thommes, Nagasawa \& Lin 2008), depending on the modelling of collisional accretion/fragmentation process.
Another possible source of dissipation is the 
 interaction with a remaining gas disk 
(Papaloizou \& Larwood 2000; Agnor \& Ward 2002; Kominami \& Ida 2002), although
this scenario too, has been criticised.
The effects of gas drag have to be
tuned somewhat, otherwise the drag can be either too strong (preventing orbit crossing to begin with) or irrelevant
(Kominami \& Ida 2002). Such tuning may, however, emerge naturally in a model where the migration of Jupiter in a
gaseous disk excites the eccentricities of the terrestrial embryos by resonance sweeping (Nagasawa, Lin \& Thommes 2005;
Thommes, Nagasawa \& Lin 2008),
thereby guaranteeing that the planet assembly occurs while the gas density is still high enough to be significant.

The first point to note is the empirical fact that the PSR~B1257+12 system occupies an even more extreme position
in the parameter space than the terrestrial system, with a similar value of $S_d$ but a significantly higher $S_c$.
This poses an interesting problem for the aforementioned resonant sweeping scenario, since there are no Jovian mass
planets in the system to trigger the planet assembly. While a population of smaller bodies or gas drag may indeed operate
in this scenario too, it cannot be triggered by the mechanism suggested by Nagasawa et al. However, a second point
to note is that our simulations do not overpredict the value of $S_d$, and can match both the pulsar planet and
terrestrial system quite easily. Thus it does not appear necessary to invoke an additional dissipation mechanism  
to explain the eccentricities, although it may be required to explain the  remaining mismatch in $S_c$.
 The reason our simulations match
$S_d$ is that the material is initially concentrated in a relatively narrow range of radii, so that the planetary eccentricities do not need to grow to the levels required to induce orbit crossing and collisions in traditional terrestrial planet
accumulation studies. Whether a similar concentration of material is a plausible starting point for our own
Solar System is a subject for future calculation.

\subsection{Further Refinements}
\label{OtherStuff}

Although the above results are very encouraging, there are still some remaining discrepancies. The simulations
can match the observations in most of the statistics proposed by Chambers, but the one remaining mismatch, in
$S_c$, suggests that there may still be some physics missing.
 In this case, there is still possibly
some additional form of dissipation required from a population of small bodies or gas drag, although it cannot
be tuned by Jovian planet migration as has been proposed for our solar system.
 We have also ignored any effects due to possible
radial migration due to torques with any remnant gas disk. This could potentially explain both the more compact
configuration and the fact that the best-fit annulus is slightly closer to the star than suggested by the
gaseous disk simulations. Tidal effects were also briefly considered but ruled out as irrelevant in this context.

\section{Conclusion}

By systematically modelling the various formation scenarios proposed for the PSR~B1257+12 planets, we have
arrived at the conclusion that the most likely origin for the observed planet properties is formation from
an expanding disk of roughly solar composition, with total angular momentum $\sim 10^{49}$ ergs\, s. This
is towards the low end of the angular momentum range of proposed scenarios, and consistent with models that
postulate an origin for the disk in material captured during supernova fallback (although significantly super
solar metallicity does not provide an outcome consistent with the observations).

We furthermore find that the models that provide the best fit are those in which the gaseous disk expands
in the popular `layered' disk formalism, and that the material that condenses out of the nebula forms planetesimals
rapidly and becomes decoupled from the gas on a timescale short compared to the viscous evolution time. This
results in the majority of the solid material being deposited in a narrow annulus. The subsequent collisional
assembly of the final planets produces planets of the right mass and approximate location, although some additional
dissipation mechanism may be necessary to explain the compactness of the observed system.

\acknowledgements BH acknowledges support by the Sloan Foundation, the 
NASA Astrobiology Society and the Space Interferometry Mission. In particular, the computing
cluster purchased with Sloan Funds made the calculations presented in this paper
considerably less painful than they would otherwise have been.

\newpage

\newpage
\plotone{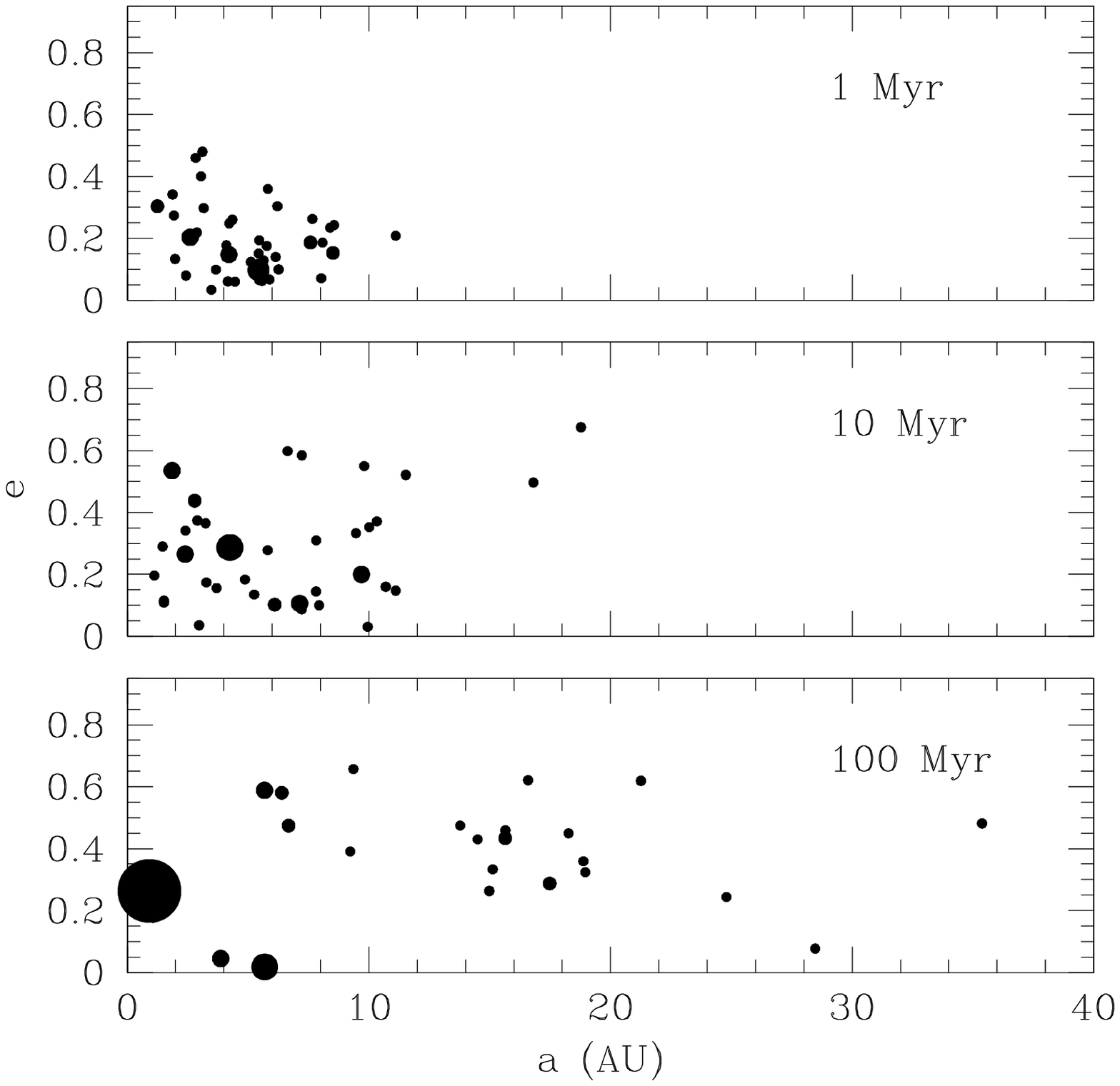}
\figcaption[f1]{The three panels show the dynamical state of the planetary embryo disk
 in the case of a fully viscous
accretion disk
with total angular momentum $J=10^{51}$~ergs\, s, at ages of 1~Myr, 10~Myr and 100~Myr from top
to bottom.
 The size of the points scales linearly with the mass
of the body, so we see that the most massive body grows to $16 M_{\oplus}$ by the end
and is located at $0.91$AU.
 \label{Snap1}}
\newpage
\plotone{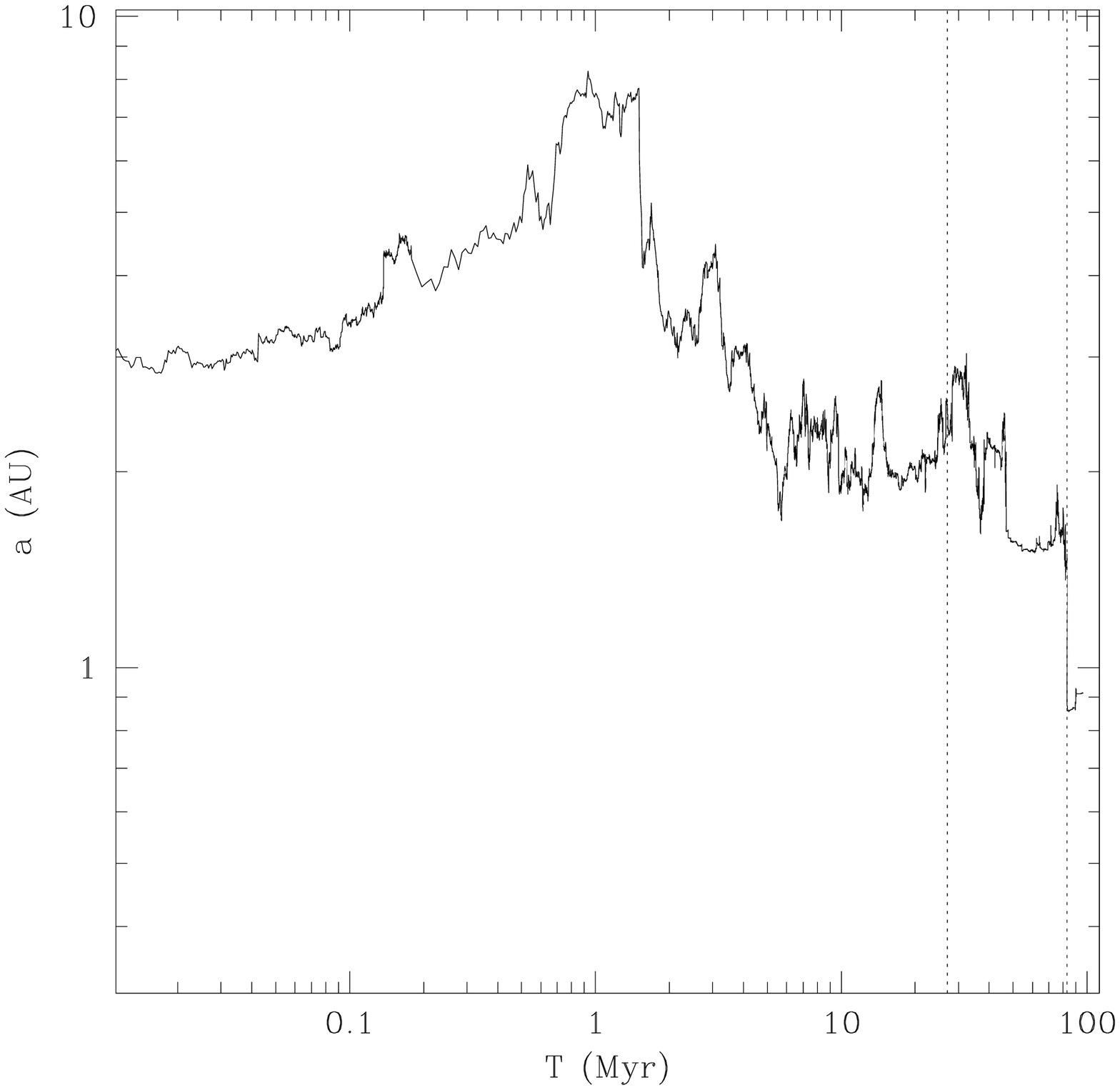}
\figcaption[f2.ps]{The radial evolution of the body that becomes the most massive one
in Figure~\ref{Snap1} is shown here. The early growth period is characterised by an outward
migration as simply part of a random walk, but as the body comes to be the most massive it scatters smaller bodies outwards
and, as a result, migrates inward. The dotted lines indicate the times at which the mass
crosses the thresholds of 5$M_{\oplus}$ and 10$M_{\oplus}$ respectively. In the latter case,
this is the result of the accretion of another large body of $6 M_{\oplus}$, which also
results in a significant change in semi-major axis and is responsible for the large
final eccentricity (e=0.263).
 \label{Massive1}}
\newpage
\plotone{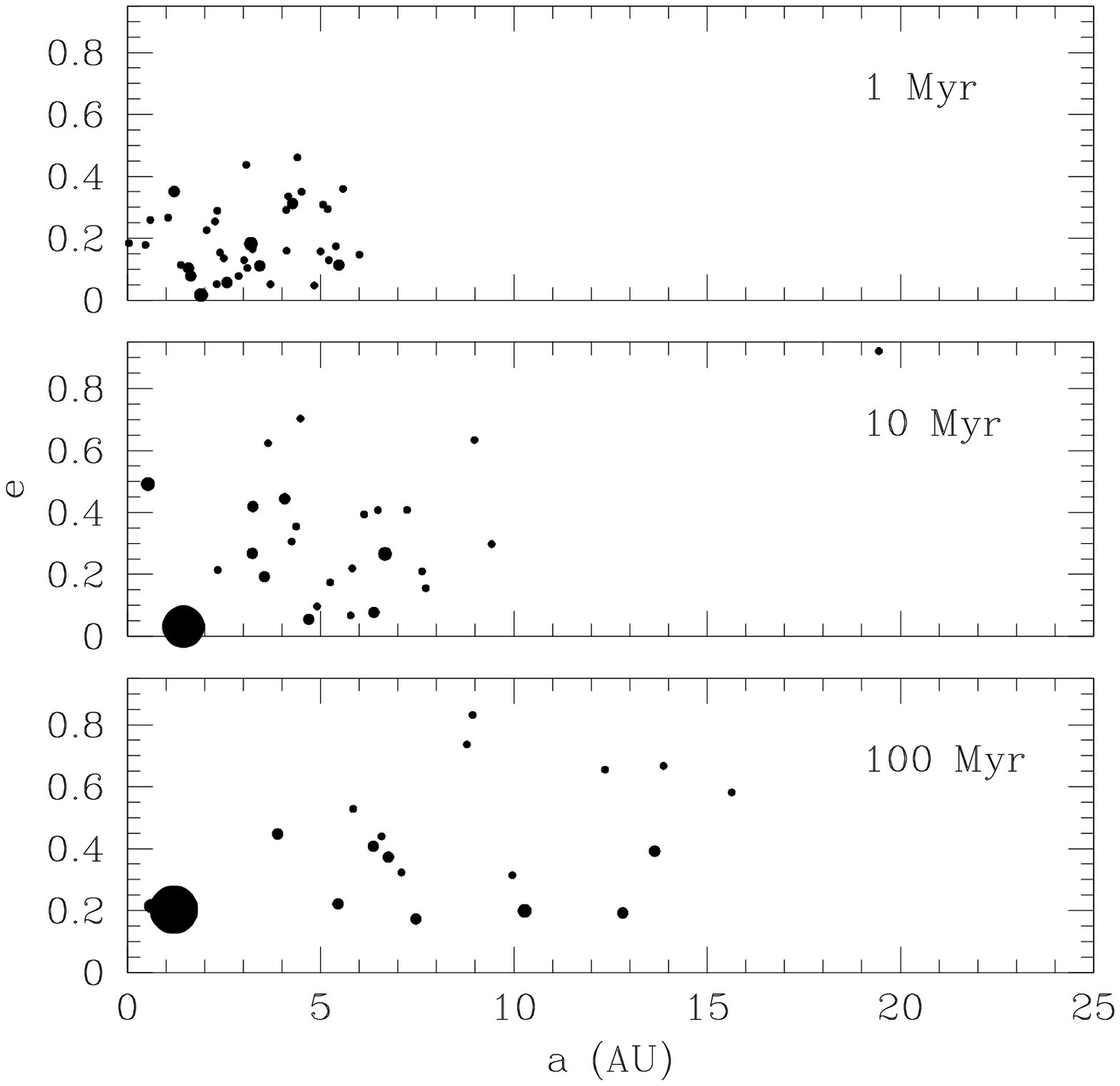}
\figcaption[f3.ps]{The three panels show the evolutionary state of the planetary
embryo population that result from a layered disk evolution of a $J=10^{51}$ergs\, s 
model, for ages of 1 Myr, 10 Myr and 100 Myr. The size of the points is linearly scaled
to represent the mass of the body. We see that the dynamics is dominated by a single
large body, $\sim 12 M_{\oplus}$, that forms earlier than in the previous case. Nevertheless,
the rapid inwards migration still leaves a remnant scattered population on scales of
tens of AU.
 \label{Snap2}}
\newpage

\plotone{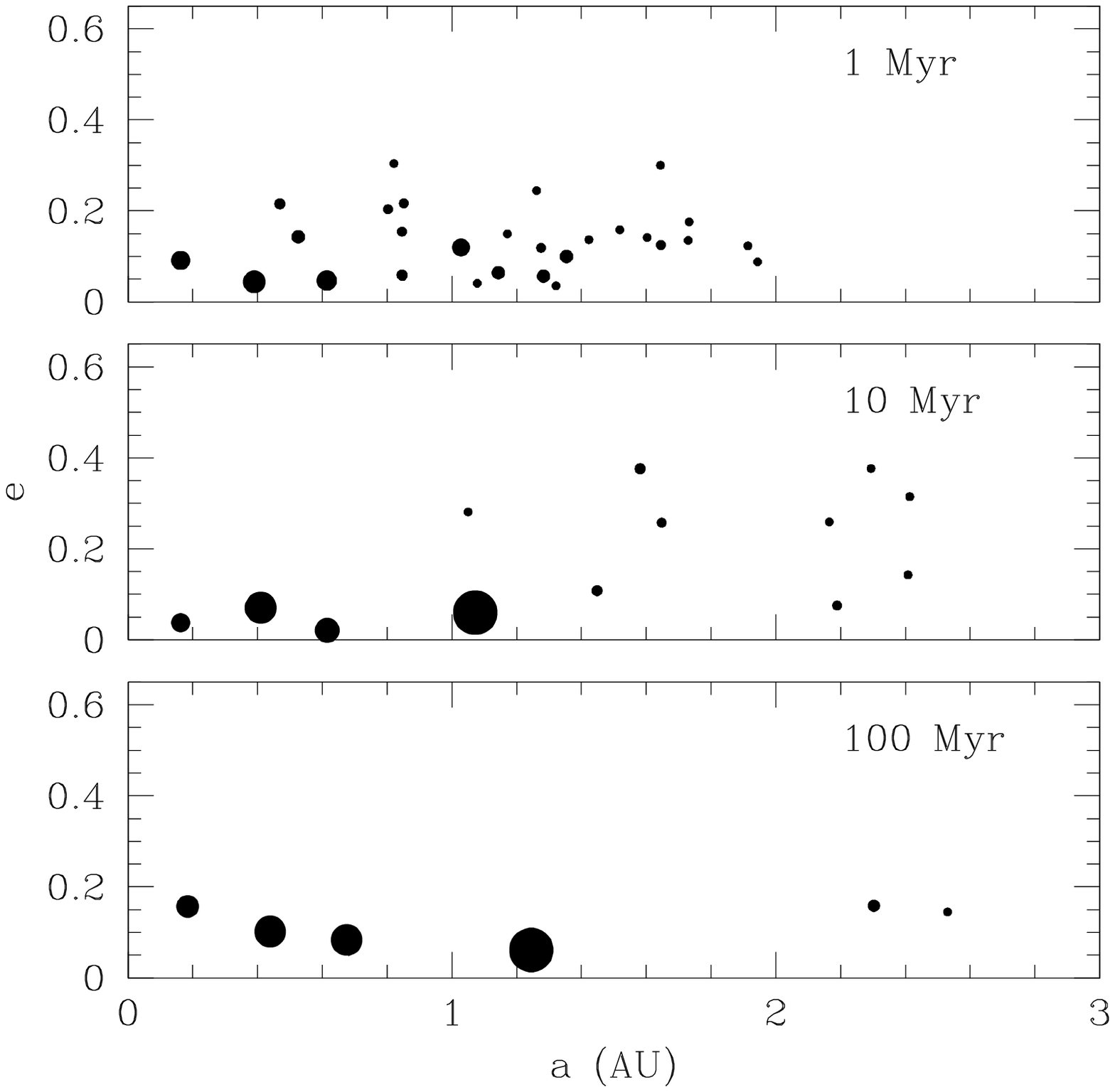}
\figcaption[f4.ps]{The panels show the dynamical evolution of the planetary embryo swarm
that results from a fully viscous disk that starts with $J=10^{49}$~ergs\, s.  Once
again the size of the points scales linearly with their mass (although we note that the 
scaling is different than that used in Figures~\ref{Snap1} and \ref{Snap2}.) We see that
the evolution is quite rapid, with the assembly of the final planetary configuration largely
complete within $10^7$ years.
\label{Snap3}}

\newpage
\plotone{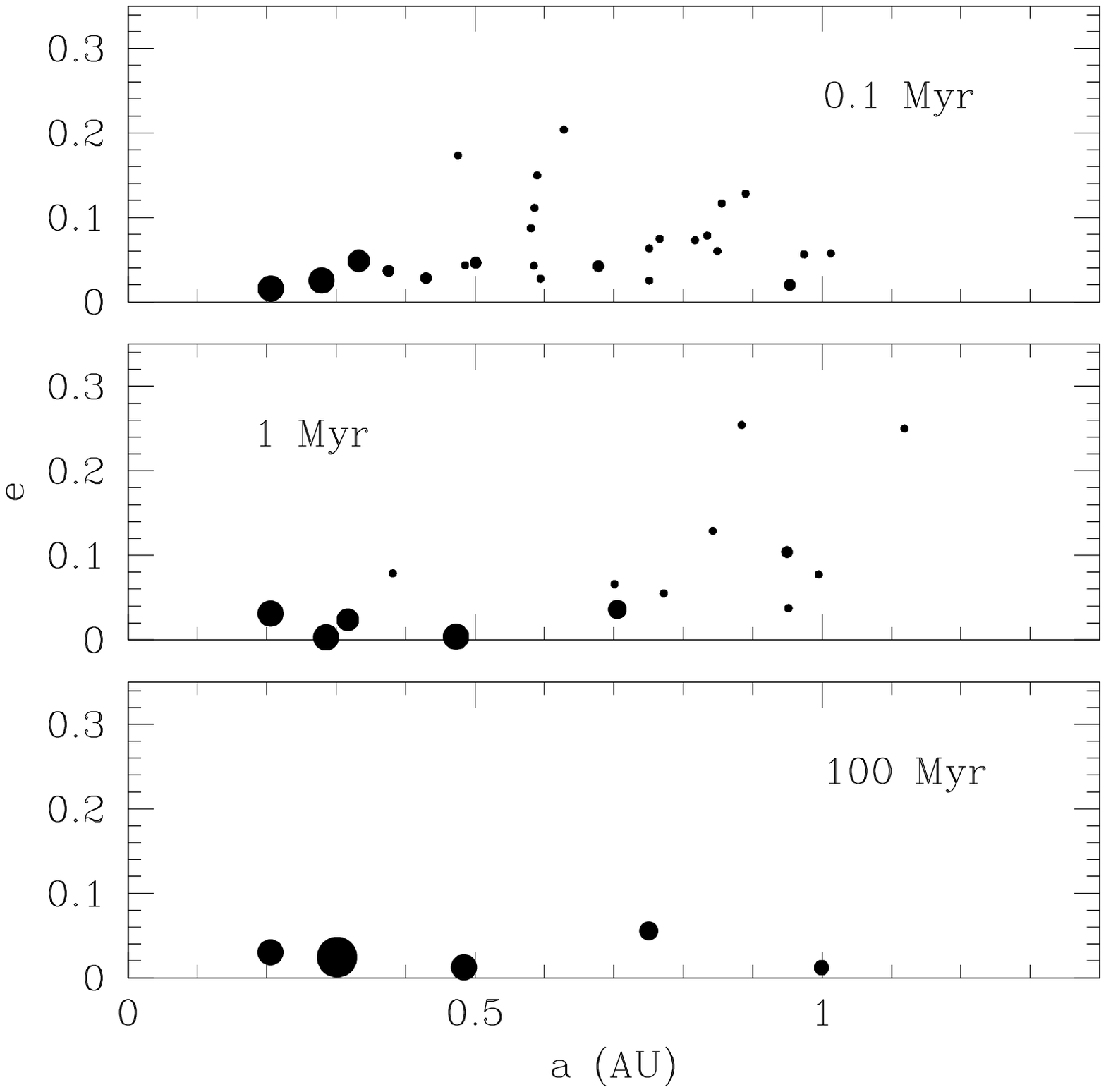}
\figcaption[f5.ps]{The panels show the dynamical evolution of the planetary embryo swarm
that results from a disk evolving via the layered accretion model  with $J=10^{49}$~ergs\, s.
Once again the size of the points scales linearly with their mass (although we note that the
scaling is different than that used in Figures~\ref{Snap1} and \ref{Snap2}.) The evolution
proceeds rapidly, especially in the inner part of the nebula, where the assembly is largely
complete within $10^5$~years. On timescales $\sim 10^7$~years,
 the region around 1~AU accretes together to form the outer two planets. An encouraging
sign is that this model produces planets of about the right mass at $\sim 0.3$--0.5~AU,
where the planets are observed. However, the final configuration still contains too many
bodies.
\label{Snap4}}

\newpage

\plotone{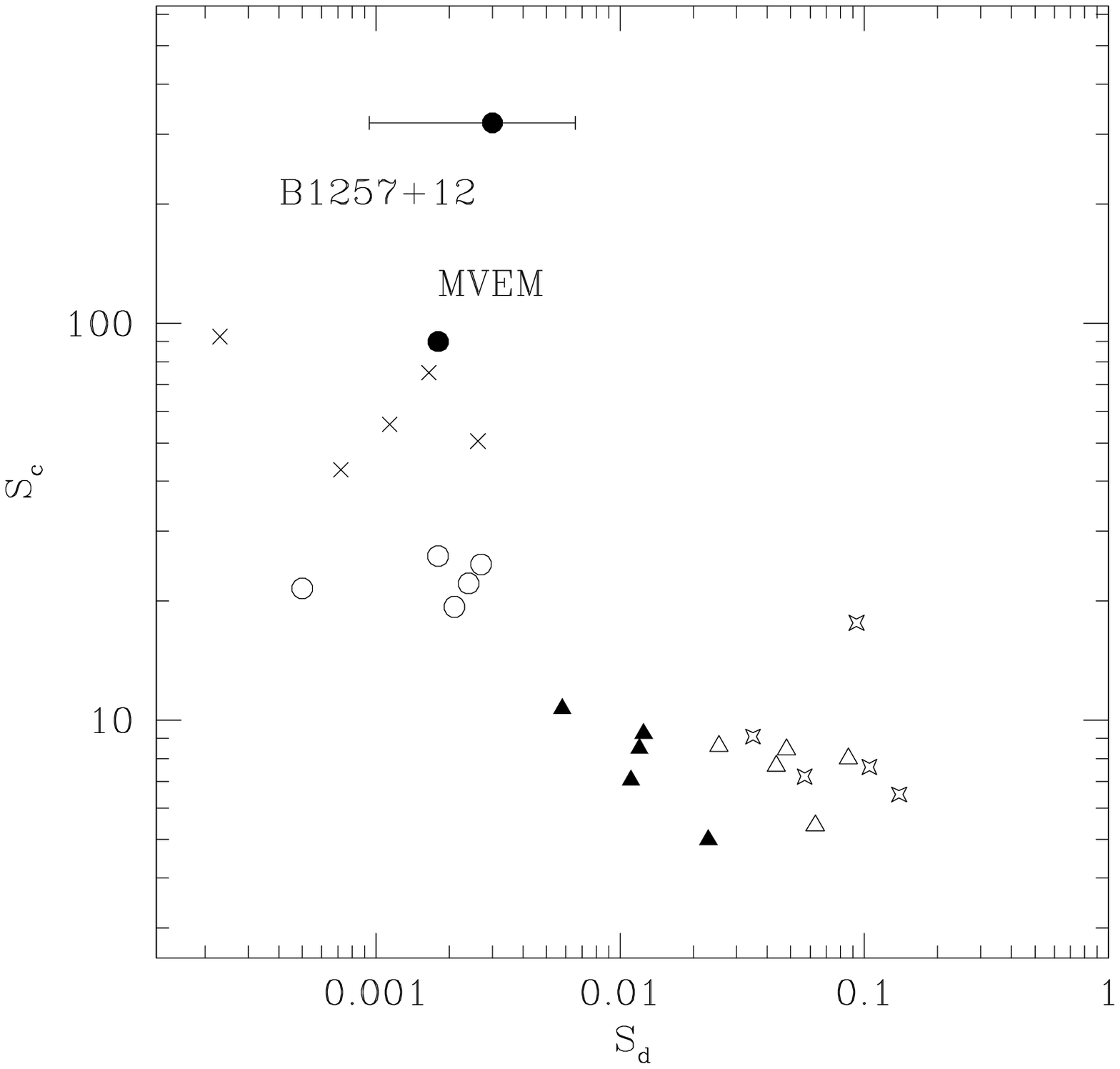}
\figcaption[f6.ps]{The filled circles indicate the location of the pulsar planet system, as
well as our own terrestrial planet system (denoted MVEM) for comparison. Also shown are the results of the
simulations for various models discussed in the text. The $J=10^{51}$ergs\, s model is denoted
as open stars (fully viscous) and triangles (layered), while the $J=10^{49}$ergs\, s model is
denoted as filled triangles (fully viscous) and open circles (layered). The crosses represent
the results from the heavy element disk models. In this metric this latter model is the best fit,
but it is severely discrepant in others, such as the mass-weighted semi-major axis $S_a$.
\label{SS2}}

\newpage

\plotone{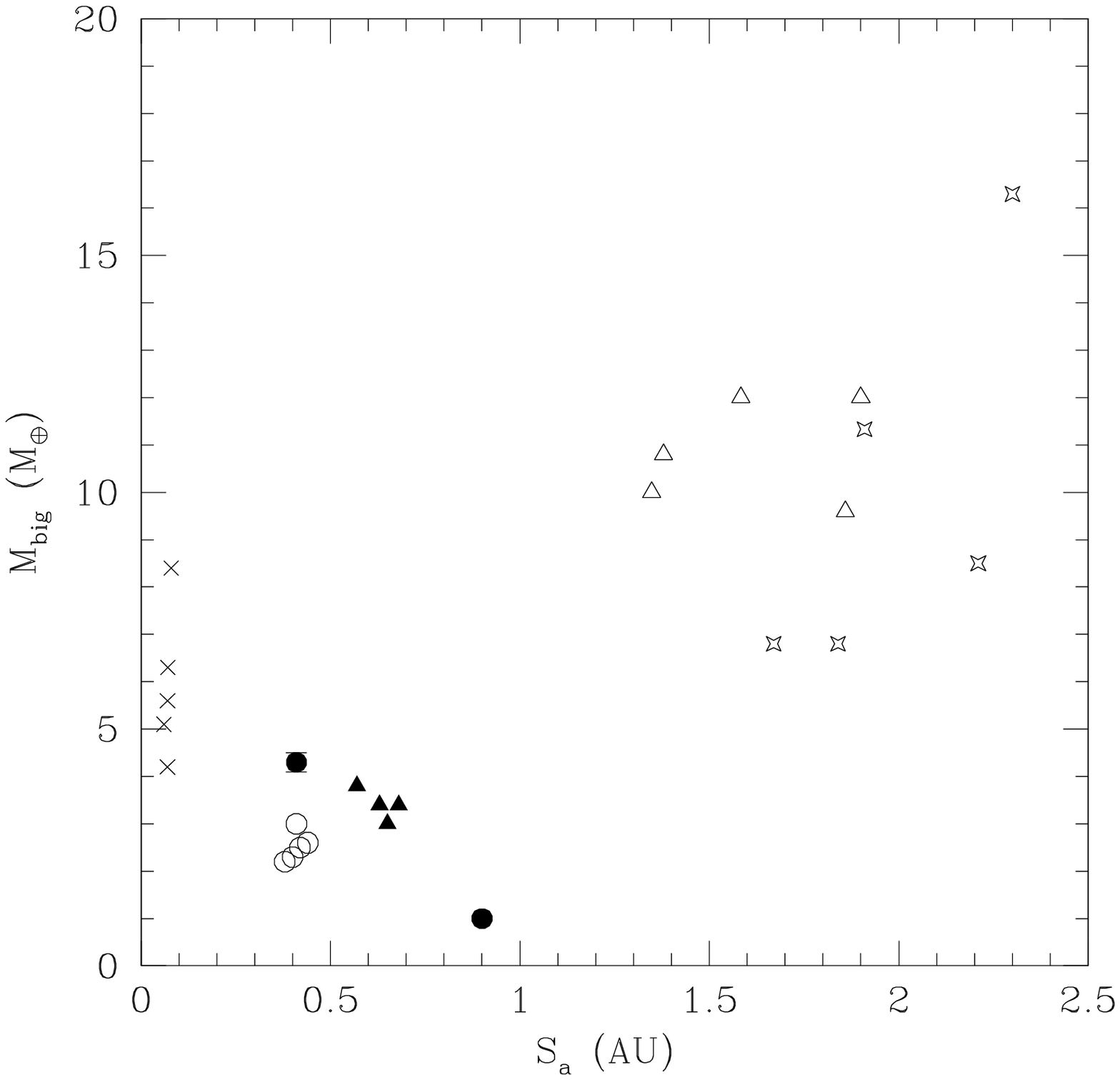}
\figcaption[f7.ps]{The symbols are the same as in Figure~\ref{SS2}, but now the comparison is
made in terms of the largest mass planet ($M_{big}$) and the mass-weighted semi-major axis $S_a$. 
We see that the lower angular momentum disks provide final configurations that are the
closest to the pulsar planet system.
\label{MA0}}

\newpage

\plotone{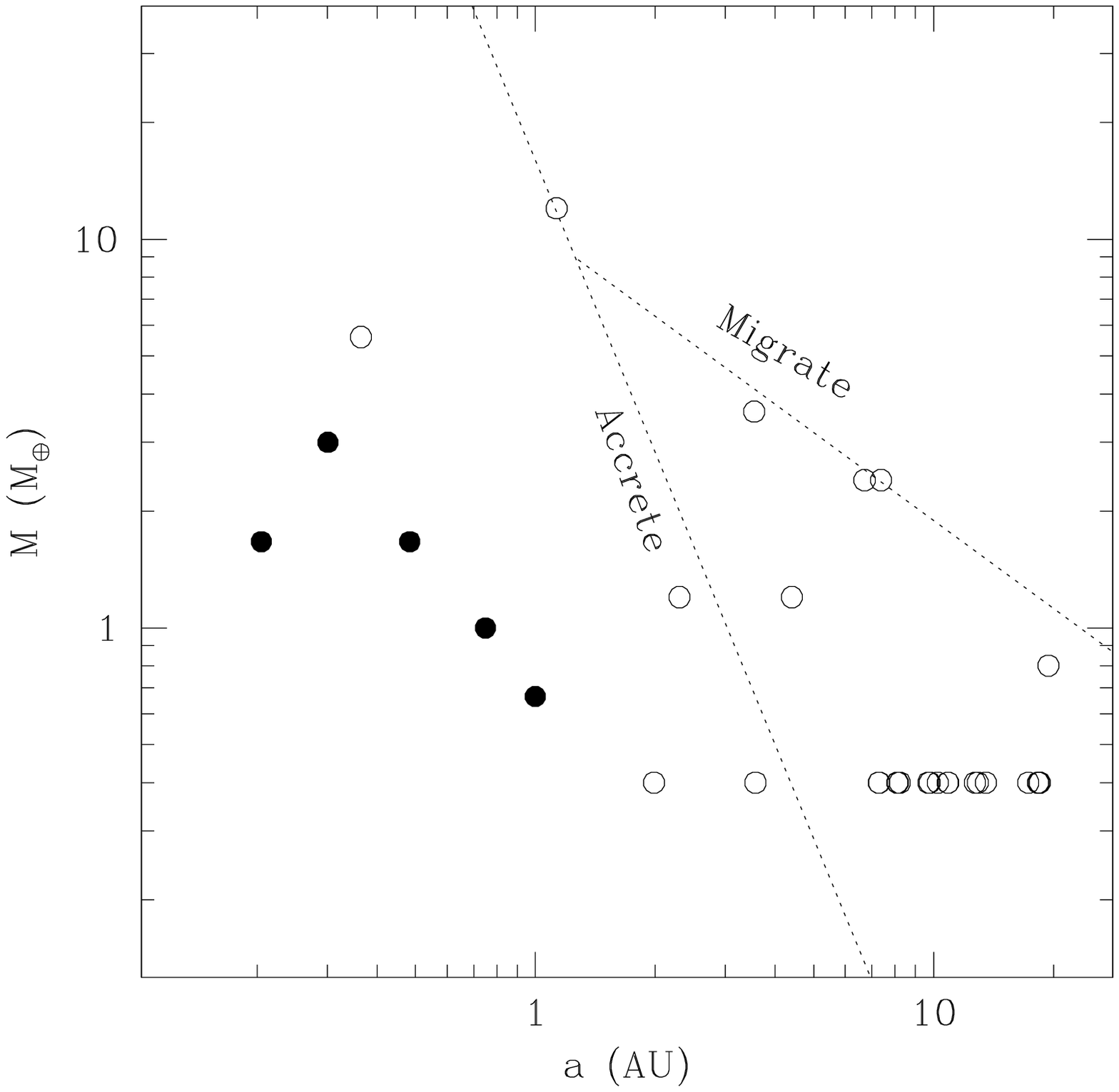}
\figcaption[f8.ps]{The solid points are the end result of the simulations shown in Figure~\ref{Snap4},
while the open points are the simulation shown in Figure~\ref{Snap2}. The dotted lines labelled `Migrate'
and `Accrete' are described in the text, and represent different dynamical regimes. In particular, bodies
that lie above the line labelled `Migrate' will move rapidly inwards, while those that lie below the line
labelled `Accrete' tend to grow in mass without changing their orbit significantly.
\label{MA3}}

\newpage

\plotone{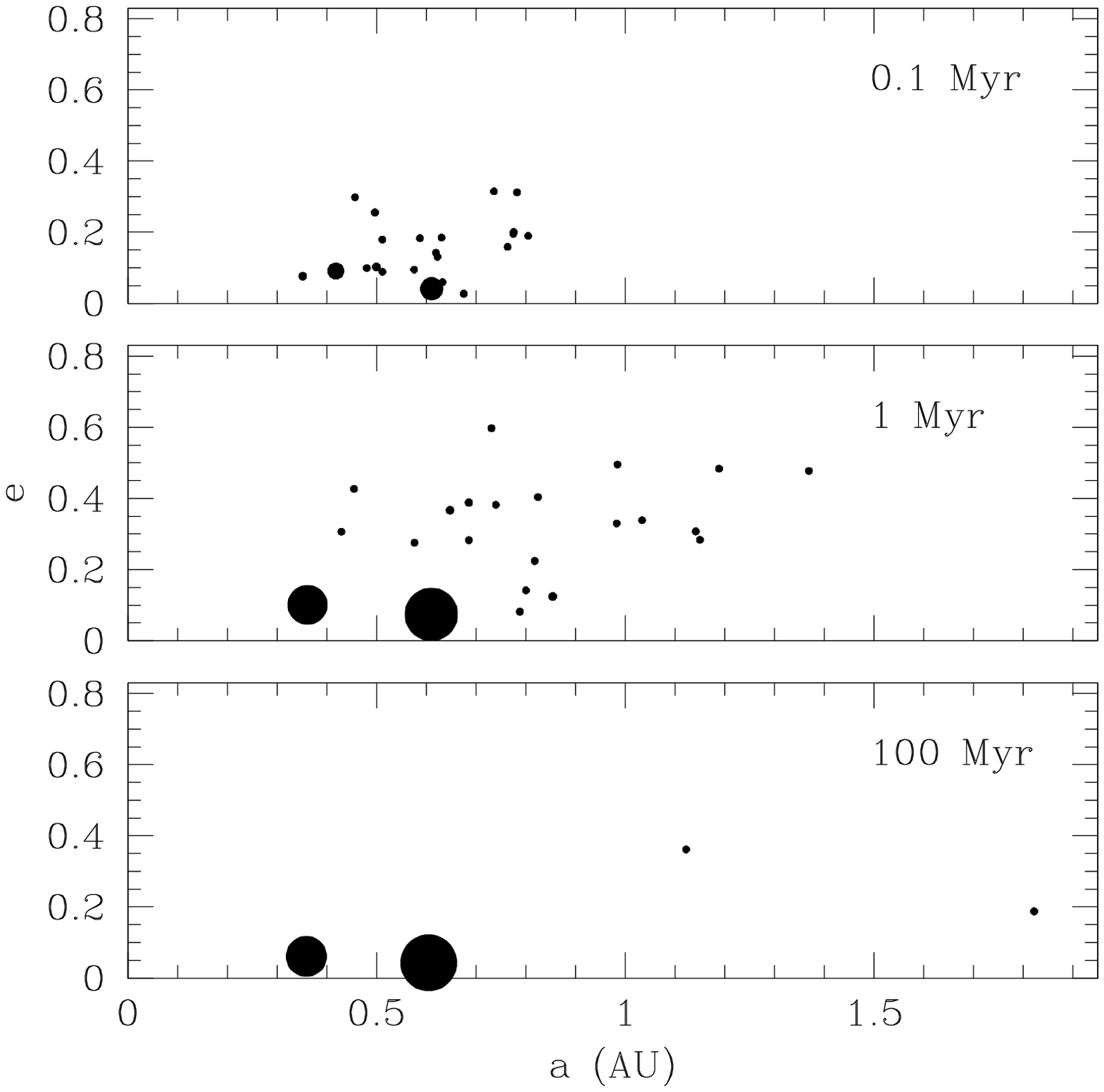}
\figcaption[f9.ps]{The panels show the dynamical evolution of the planetary embryo swarm
that results from an initial distribution confined to an annulus between 0.4--0.6~AU. This
is what we expect to result 
 from rapid sedimentation out of a layered disk with $J=10^{49}$ergs\, s.
The size of the points scale linearly with the mass. We see that the evolution is
quite rapid in this case, with the final configuration in place by 1 Myr. The
rapidity of assembly is 
 because all the mass is concentrated in a narrow region to begin
with. This scenario is the most encouraging for reproducing the pulsar planets, as the number, location and masses of the planets
all resemble the observations.
\label{Snap5}}

\newpage

\plotone{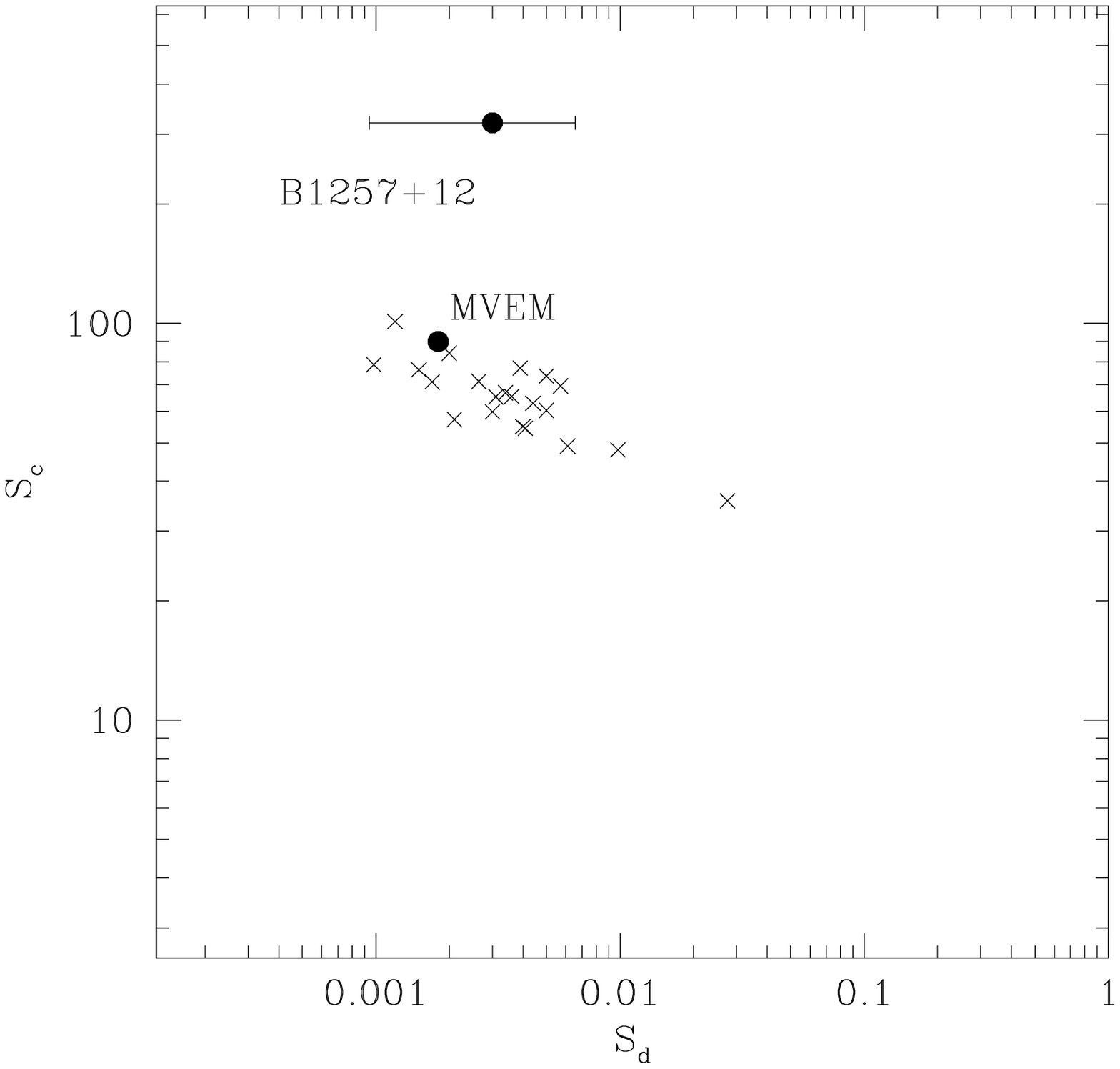}
\figcaption[f10.ps]{The filled circles indicate the location of the pulsar planet system, as
well as our own terrestrial planet system for comparison. Also shown, as crosses, are 
several realisations of the scenario in which the planetary embryos are started in an annulus
between 0.4 and 0.6~AU. We see this is in much better agreement than the results of the
simulations in Figure~\ref{SS2}. 
\label{SS3}}

\newpage

\plotone{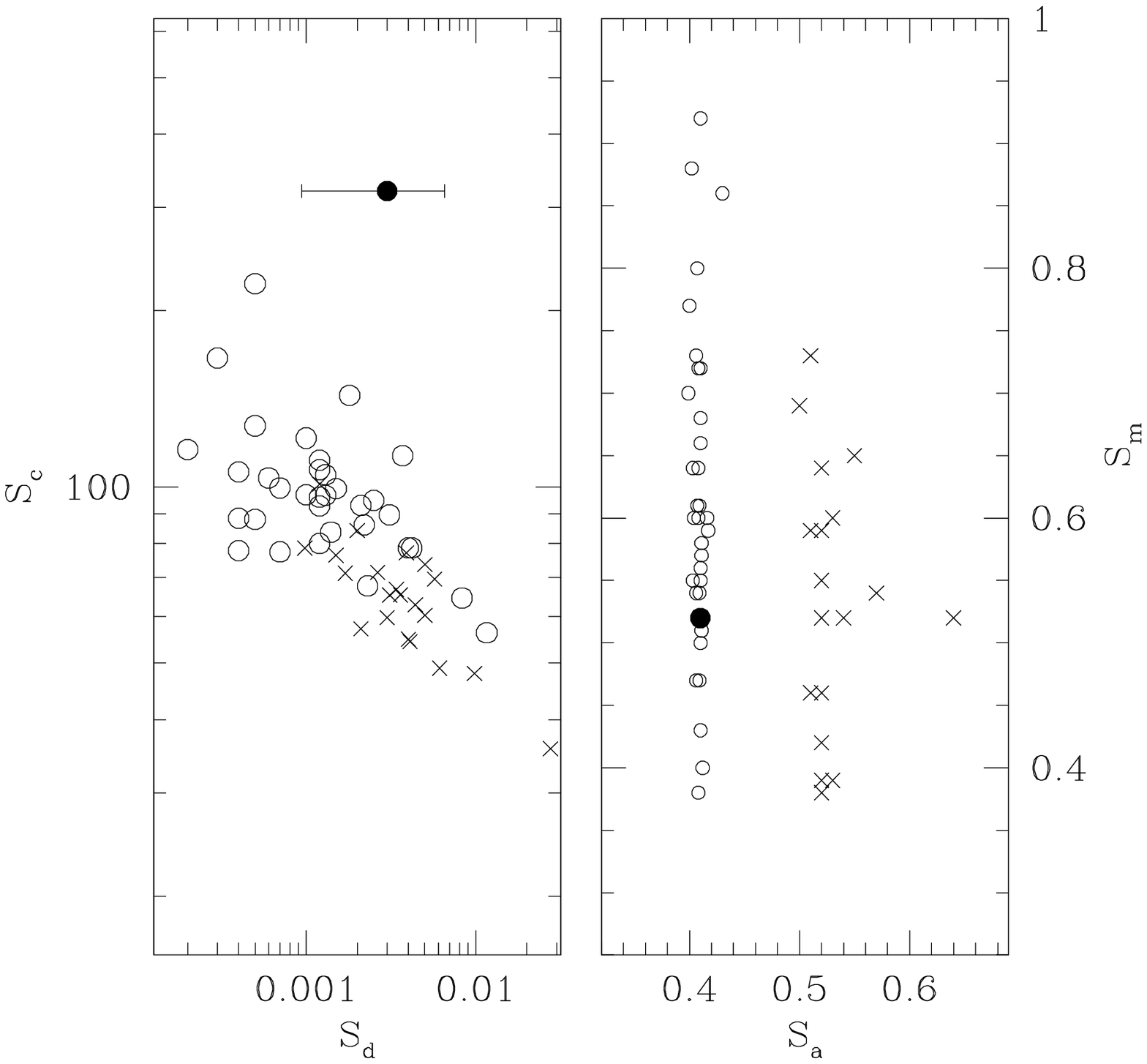}
\figcaption[f11.ps]{The filled circle indicates the location of the pulsar planet system
(we have omitted the solar system planets this time). The crosses are the same results as
shown in Figure~\ref{SS3}, while the open circles are the results for the simulations
that start with a narrow ring between $0.35$--$0.45$~AU.
\label{SS4}}

\newpage
\plotone{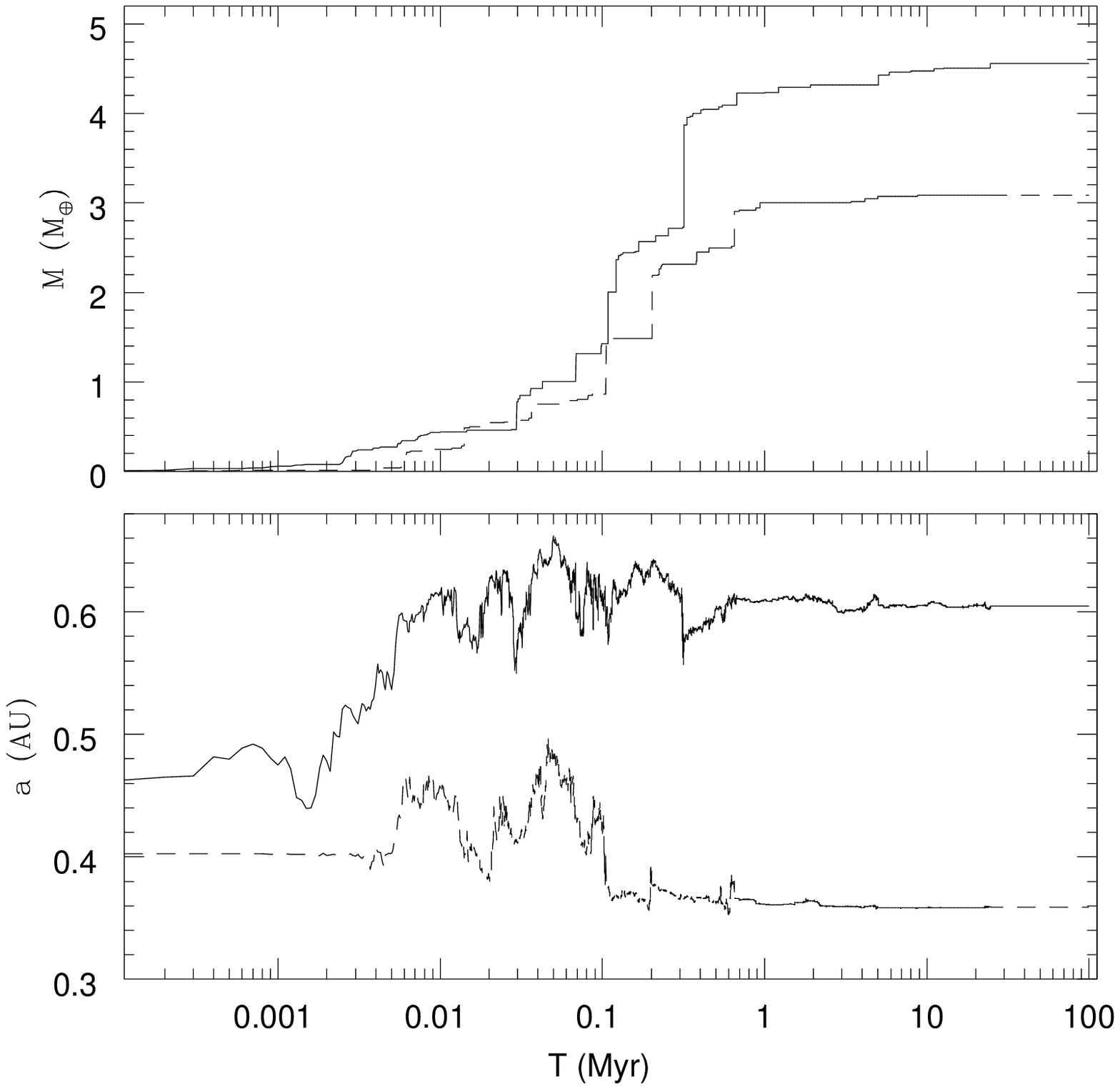}
\figcaption[f12.ps]{The upper panel shows the assembly history of the two large surviving
planets in the simulation shown in Figure~\ref{Snap5}. We see that the quickest period of
growth occurs between 0.1 and 1 Myr, including several large jumps in mass corresponding to
major impacts. The lower panel shows the orbital history of the same two bodies. The interaction
through the process of planetesimal scattering drives them apart during the assembly process.
\label{Assemble}}

\newpage
\plotone{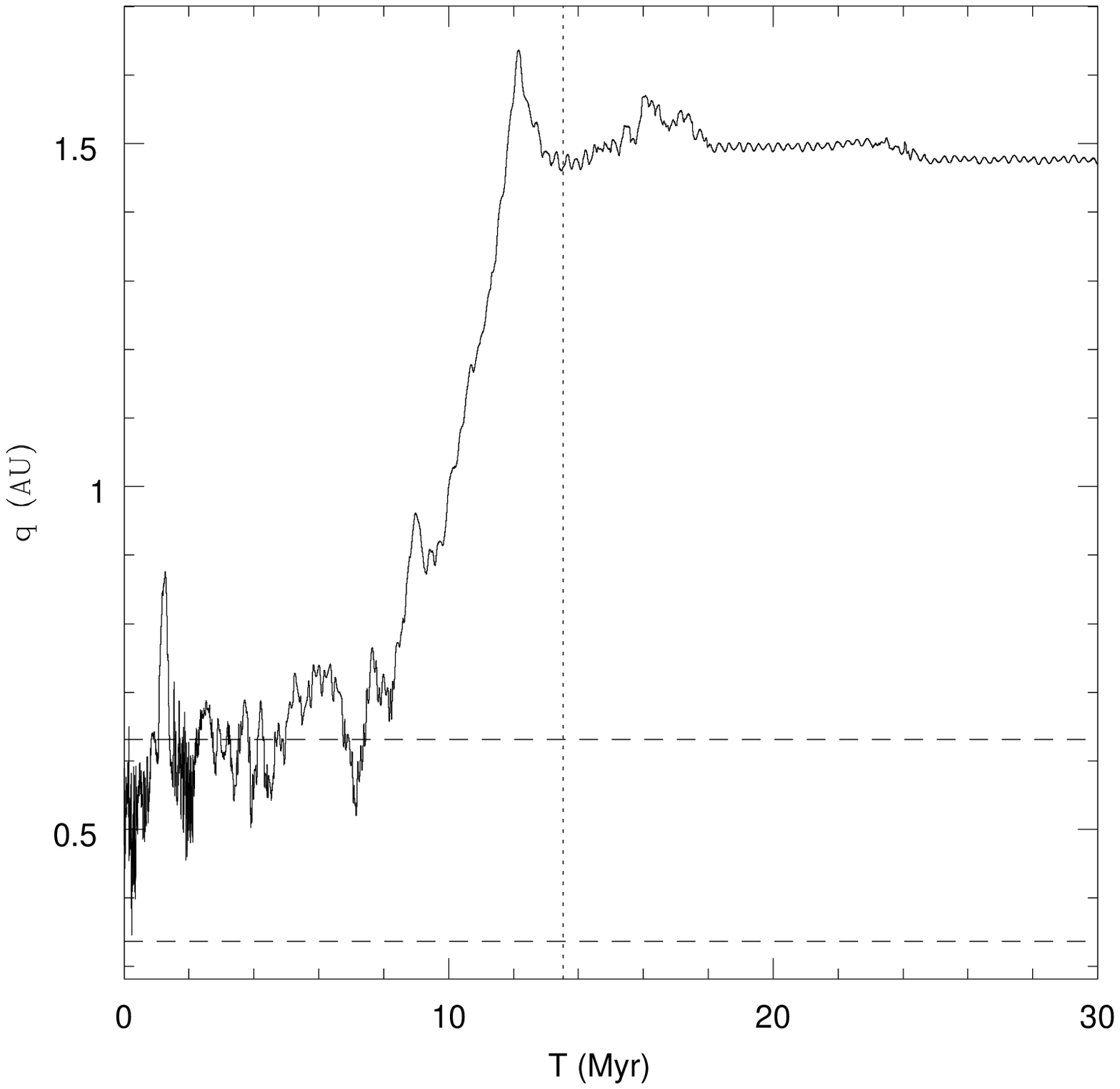}
\figcaption[f13.ps]{The solid line shows the evolution of the periastron of the outermost surviving
body in Figure~\ref{Snap5}. The dashed lines show the region in which close scattering by the
large bodies in the system is possible. The original decoupling from the strong scattering region
is driven by a gravitational interaction with another small, scattered body. When this intermediary
is eventually ejected from the system (marked by the vertical dotted line), the periastron ceases
to evolve significantly, leaving a remnant that is dynamically decoupled from the interior system.
\label{Ast232}}

\begin{deluxetable}{lccc}
\tablecolumns{4}
\tablewidth{0pc}
\tablecaption{Physical Parameters for the PSRB1257+12 planetary system, taken from Konacki \& Wolszczan (2003).
Figures in parenthesis indicate the formal 1$\sigma$ error bars. Note that although the inclination relative to the
observer has two possible values, the relative inclination is essentially identical.
 \label{Params}}
\tablehead{
\colhead{Parameter}    &  \colhead{Planet~A} & \colhead{Planet~B} & \colhead{Planet~C} 
 }
\startdata
Planet Semi-major axis (AU) & 0.19 & 0.36 & 0.46 \\
Orbital Period (days)       & 25.262(3) & 66.5419(1) & 98.2114(2) \\
Planet Mass ($M_{\oplus}$)  & 0.020(2) & 4.3(2) & 3.9(2) \\
Eccentricity                & 0.0 & 0.0186(2) & 0.0252(2) \\
Inclination (degrees)       & $\cdots$ & 53(4) & 47(3) \\
                            & $\cdots$ & 127(4) & 133(3) \\
\enddata
\end{deluxetable}

\begin{deluxetable}{lcccccccc} 
\tablecolumns{9} 
\tablewidth{0pc} 
\tablecaption{Planetary System Statistics for simulations of various scenarios. ``LD'' refers to Large Disk;
``SD'' to Small Disk; ``HD'' to Heavy Disk, and ``V'' or ``L'' refers to either a fully viscous evolution
or a layered disk evolution. \label{CompTab}} 
\tablehead{ 
\colhead{Name}    &  \colhead{N} & \colhead{$S_m$} & \colhead{$S_a$} & \colhead{$S_s$} & \colhead{$S_d$} & \colhead{$S_c$} &
\colhead{$M_{big}$} & \colhead{ $a_{big}$} \\
\multicolumn{3}{c}{} & \colhead{(AU)} & \multicolumn{3}{c}{} & \colhead{ ($M_{\oplus}$)} & \colhead{(AU)} }
\startdata 
\sidehead{Simulations}
 LDV1 & 6 & 0.49 & 2.30 & 17.1 & 0.139 & 6.5 & 16.3 & 0.91 \\
 LDV2 & 4 & 0.37 & 2.22 & 29.0 & 0.035 & 9.1 & 8.5 & 1.99 \\
 LDV3 & 4 & 0.40 & 1.84 & 32.2 & 0.105 & 7.6 & 6.8 & 0.52 \\
 LDV4 & 6 & 0.38 & 1.67 & 21.7 & 0.057 & 7.2 & 6.8 & 1.96 \\
 LDV5 & 7 & 0.74 & 1.91 & 14.8 & 0.093 & 17.6 & 11.3 & 1.16 \\
 LDL1 & 7 & 0.55 & 1.90 & 18.2 & 0.086 & 8.0 & 12.0 & 1.21 \\
 LDL2 & 7 & 0.49 & 1.18 & 17.9 & 0.048 & 8.5 & 12.0 & 1.13 \\
 LDL3 & 4 & 0.56 & 1.38 & 32.7 & 0.025 & 8.6 & 10.8 & 1.90 \\
 LDL4 & 6 & 0.52 & 1.35 & 22.8 & 0.044 & 7.7 & 10.0 & 0.87 \\
 LDL5 & 5 & 0.36 & 1.86 & 20.6 & 0.063 & 5.4 & 9.6 & 0.38 \\
 SDV1 & 6 & 0.35 & 0.68 & 26.5 & 0.0058 & 10.7 & 3.4 & 1.25 \\
 SDV2 & 7 & 0.29 & 0.65 & 24.7 & 0.0111 & 7.1 & 3.0 & 1.06 \\
 SDV3 & 9 & 0.33 & 0.63 & 19.5 & 0.0125 & 9.3 & 3.4 & 0.24 \\
 SDV4 & 9 & 0.36 & 0.63 & 25.9 & 0.0120 & 8.5 & 3.4 & 0.72 \\
 SDV5 & 8 & 0.38 & 0.57 & 21.1 & 0.0230 & 5.0 & 3.8 & 0.80 \\
 SDL1 & 5 & 0.37 & 0.41 & 25.4 & 0.0005 & 21.5 & 3.0 & 0.30 \\
 SDL2 & 5 & 0.27 & 0.38 & 29.4 & 0.0027 & 24.7 & 2.2 & 0.69 \\
 SDL3 & 5 & 0.29 & 0.40 & 25.2 & 0.0021 & 19.3 & 2.3 & 0.22 \\
 SDL4 & 5 & 0.31 & 0.42 & 25.2 & 0.0024 & 22.1 & 2.5 & 0.29 \\
 SDL5 & 6 & 0.33 & 0.44 & 22.3 & 0.0018 & 25.9 & 2.6 & 0.36 \\
 HD1 & 3 & 0.42 & 0.08 & 20.9 & 0.0026 & 50.5 & 8.4 & 0.09 \\
 HD2 & 2 & 0.64 & 0.06 & 41.3 & 0.0007 & 42.8 & 5.1 & 0.09 \\
 HD3 & 2 & 0.56 & 0.07 & 27.5 & 0.0002 & 92.6 & 5.6 & 0.09 \\
 HD4 & 3 & 0.42 & 0.07 & 24.3 & 0.0011 & 55.6 & 4.2 & 0.07 \\
 HD5 & 3 & 0.63 & 0.07 & 26.2 & 0.0017 & 75.1 & 6.3 & 0.07 \\
\sidehead{Observed Systems}
B1257+12 & 3 & 0.52 & 0.41 & 28.0 &  0.0037(28) & 320 & 4.3(2) & 0.36\\
MVEM     & 4 & 0.51 & 0.90 & 37.7 & 0.0018 &  90 & 1.0 & 1.0 \\
\enddata 
\end{deluxetable}

\begin{deluxetable}{lcccccccc}
\tablecolumns{9}
\tablewidth{0pc}
\tablecaption{Planetary System Statistics compared to annulus models. \label{CompTab2}}
\tablehead{
\colhead{Name}    &  \colhead{N} & \colhead{$S_m$} & \colhead{$S_a$} & \colhead{$S_s$} & \colhead{$S_d$} & \colhead{$S_c$} &\colhead{$M_{big}$} & \colhead{ $a_{big}$} \\
\multicolumn{3}{c}{} & \colhead{(AU)} & \multicolumn{3}{c}{} & \colhead{ ($M_{\oplus}$)} & \colhead{(AU)} }
\startdata
\sidehead{0.4--0.6~AU}
 FLR1 & 3 & 0.69 & 0.50 & 29.0  & 0.0039 & 77.1 & 5.7 & 0.53\\
 FLR2 & 3 & 0.73 & 0.51 & 42.6  & 0.0057 & 69.5 & 6.1 & 0.42 \\
 FLR3 & 4 & 0.55 & 0.52 & 28.6  & 0.0040 & 54.9 & 4.6 & 0.38 \\
 FLR4 & 3 & 0.46 & 0.51 & 25.0  & 0.0036 & 65.4 & 3.7 & 0.50 \\
 FLR5 & 4 & 0.60 & 0.53 & 28.5  & 0.0050 & 73.7 & 4.8 & 0.53 \\
 FLR6 & 3 & 0.46 & 0.52 & 25.8  & 0.0050 & 60.3 & 3.7 & 0.38\\
 FLR7 & 4 & 0.65 & 0.55 & 27.1  & 0.0020 & 84.2 & 5.6 & 0.57 \\
 FLR8 & 6 & 0.52 & 0.64 & 31.7  & 0.0275 & 35.7 & 4.5 & 0.44 \\
 FLR9 & 4 & 0.54 & 0.57 & 25.1  & 0.0017 & 71.2 & 4.6 & 0.42 \\
FLR10 & 4 & 0.59 & 0.51 & 32.9 & 0.0026 & 71.4 & 4.6 & 0.61 \\
FLR11 & 4 & 0.38 & 0.52 & 22.1 & 0.0010 & 78.6 & 2.9 & 0.38 \\
FLR21 & 3 & 0.52 & 0.52 & 30.5 & 0.0031 & 65.3 & 4.1 & 0.39 \\
FLR22 & 3 & 0.64 & 0.52 & 31.9 & 0.0044 & 62.9 & 5.0 & 0.50 \\
FLR23 & 4 & 0.39 & 0.53 & 25.7 & 0.0030 & 59.8 & 3.0 & 0.37 \\
FLR24 & 3 & 0.59 & 0.52 & 33.9 & 0.0015 & 76.4 & 4.6 & 0.40 \\
FLR25 & 6 & 0.52 & 0.54 & 26.4 & 0.0098 & 48.0 & 4.0 & 0.39 \\
FLR26 & 4 & 0.46 & 0.52 & 26.7 & 0.0034 & 66.8 & 3.6 & 0.39 \\
FLR27 & 3 & 0.39 & 0.52 & 25.4 & 0.0021 & 57.2 & 3.0 & 0.36 \\
FLR28 & 6 & 0.42 & 0.52 & 25.8 & 0.0061 & 49.0 & 3.3 & 0.61 \\
\sidehead{$0.35$--$0.45$AU}
FLR15 & 2 & 0.77 & 0.401 & 25.4 & 0.0003 & 166 & 6.4 & 0.36 \\
FLR16 & 5 & 0.86 & 0.433 & 19.8 & 0.0116 & 56.3 & 7.2 & 0.36 \\
FLR17 & 3 & 0.56 & 0.411 & 24.8 & 0.0004 & 88.4 & 4.7 & 0.38 \\
FLR18 & 2 & 0.66 & 0.408 & 30.0 & 0.0012 & 92.8 & 5.5 & 0.35 \\
FLR19 & 3 & 0.92 & 0.412 & 28.9 & 0.0037 & 113 & 7.7 & 0.37 \\
FLR20 & 3 & 0.68 & 0.411 & 26.5 & 0.0012 & 111 & 5.7 & 0.43 \\
FLR29 & 3 & 0.64 & 0.408 & 31.5 & 0.0005 & 88 & 4.9 & 0.33 \\
FLR30 & 3 & 0.60 & 0.404 & 22.8 & 0.0004 & 106 & 4.7 & 0.43 \\
FLR31 & 3 & 0.58 & 0.411 & 23.8 & 0.0013 & 96.7 & 4.5 & 0.42 \\
FLR32 & 4 & 0.55 & 0.410 & 47.9 & 0.0007 & 99.6 & 4.3 & 0.48 \\
FLR33 & 4 & 0.54 & 0.406 & 28.9 & 0.0015 & 99.3 & 4.2 & 0.32 \\
FLR34 & 3 & 0.38 & 0.408 & 18.7 & 0.0012 & 107 & 3.0 & 0.42 \\
FLR35 & 3 & 0.47 & 0.406 & 22.3 & 0.0004 & 77.8 & 3.7 & 0.30 \\
FLR36 & 3 & 0.61 & 0.409 & 27.6 & 0.0021 & 92.9 & 4.7 & 0.43 \\
FLR37 & 3 & 0.51 & 0.411 & 37.2 & 0.0013 & 105 & 4.0 & 0.48 \\
FLR38 & 4 & 0.64 & 0.403 & 31.9 & 0.0025 & 94.8 & 5.0 & 0.36 \\
FLR39 & 3 & 0.72 & 0.410 & 35.1 & 0.0031 & 89.5 & 5.6 & 0.45 \\
FLR40 & 3 & 0.57 & 0.411 & 32.0 & 0.0014 & 83.7 & 4.4 & 0.32 \\
FLR41 & 3 & 0.61 & 0.407 & 23.0 & 0.0010 & 96.9 & 4.7 & 0.33 \\
FLR42 & 3 & 0.88 & 0.402 & 35.1 & 0.0018 & 143.3 & 6.9 & 0.37 \\
FLR43 & 2 & 0.73 & 0.406 & 21.3 & 0.0005 & 222.4 & 5.7 & 0.37 \\
FLR44 & 3 & 0.55 & 0.403 & 27.5 & 0.0006 & 103.6 & 4.2 & 0.33 \\
FLR45 & 4 & 0.60 & 0.416 & 26.3 & 0.0023 & 67.7 & 4.7 & 0.32 \\
FLR46 & 2 & 0.70 & 0.399 & 28.4 & 0.0002 & 115.8 & 5.4 & 0.34 \\
FLR47 & 2 & 0.80 & 0.407 & 30.8 & 0.0005 & 127.1 & 6.2 & 0.36 \\
FLR48 & 3 & 0.40 & 0.412 & 22.3 & 0.0042 & 78.6 & 3.1 & 0.31 \\
FLR49 & 4 & 0.60 & 0.408 & 22.9 & 0.0010 & 121.1 & 4.6 & 0.45 \\
FLR50 & 3 & 0.43 & 0.410 & 21.4 & 0.0012 & 96.0 & 3.4 & 0.51 \\
FLR51 & 3 & 0.47 & 0.409 & 23.2 & 0.0012 & 80.0 & 3.6 & 0.41 \\
FLR52 & 3 & 0.72 & 0.408 & 34.8 & 0.0083 & 64.6 & 5.6 & 0.40 \\
FLR53 & 3 & 0.54 & 0.409 & 41.4 & 0.0007 & 77.4 & 4.2 & 0.31 \\
FLR54 & 3 & 0.50 & 0.410 & 23.3 & 0.0022 & 86.1 & 3.9 & 0.39 \\
FLR55 & 3 & 0.59 & 0.417 & 37.3 & 0.0040 & 78.6 & 4.6 & 0.33 \\
\sidehead{Observed Systems}
B1257+12 & 3 & 0.52 & 0.408 & 28.0 & 0.0037(28) & 320 & 4.3(2) & 0.36\\
MVEM     & 4 & 0.51 & 0.898 & 37.7 & 0.0018 &  90 & 1.0 & 1.0 \\
\enddata
\end{deluxetable}

\end{document}